\documentclass[aps,prb,superscriptaddress,twocolumn,10pt]{revtex4-1}
\usepackage{graphicx}
\usepackage{wasysym}
\usepackage{listings}
\usepackage{color}
\usepackage{booktabs}
\usepackage{amssymb}

\begin{document}


\title{Area-selective deposition and B $\delta$-doping of Si(100) with BCl$_{3}$}

\author{K.J. Dwyer}
\thanks{These two authors contributed equally}
\affiliation{Department of Physics, University of Maryland, College Park, MD 20742, USA.}
\author{S. Baek}
\thanks{These two authors contributed equally}

\affiliation{Department of Physics, University of Maryland, College Park, MD 20742, USA.}
\author{Azadeh Farzaneh}
\affiliation{Department of Materials Science and Engineering, University of Maryland, College Park, MD 20742, USA.}
\author{Michael Dreyer}
\affiliation{Department of Physics, University of Maryland, College Park, MD 20742, USA.}
\author{J.R. Williams}
\affiliation{Department of Physics, University of Maryland, College Park, MD 20742, USA.}
\author{R.E. Butera}
\email{rbutera@lps.umd.edu}
\affiliation{Laboratory for Physical Sciences, 8050 Greenmead Drive, College Park, MD 20740, USA.}
\date{\today}

\begin{abstract}
B-doped $\delta$-layers were fabricated in Si(100) using BCl$_{3}$ as a dopant precursor in ultrahigh vacuum. BCl$_{3}$ adsorbed readily at room temperature, as revealed by scanning tunneling microscopy (STM) imaging. Annealing at elevated temperatures facilitated B incorporation into the Si substrate. Secondary ion mass spectrometry (SIMS) depth profiling demonstrated a peak B concentration $>$~1.2(1)~$\times$~10$^{21}$~cm$^{-3}$ with a total areal dose of 1.85(1)~$\times$~10$^{14}$~cm$^{-2}$ resulting from a 30~L BCl$_{3}$ dose at 150~$^{\circ}$C. Hall bar measurements of a similar sample were performed at 3.0~K revealing a sheet resistance of $R_{\mathrm{s}}$ = 1.91~k$\Omega\square^{-1}$, a hole concentration of $n$ = 1.90~$\times$~10$^{14}$~cm$^{-2}$ and a hole mobility of $\mu$ = 38.0~cm$^{2}$V$^{-1}$s$^{-1}$ without performing an incorporation anneal. Further, the conductivity of several B-doped $\delta$-layers showed a log dependence on temperature suggestive of a two-dimensional system. Selective-area deposition of BCl$_{3}$ was also demonstrated using both H- and Cl-based monatomic resists. In comparison to a dosed area on bare Si, adsorption selectivity ratios for H and Cl resists were determined by SIMS to be 310(10):1 and 1529(5):1, respectively, further validating the use of BCl$_{3}$ as a dopant precursor for atomic precision fabrication of acceptor-doped devices in Si.
\end{abstract}

\maketitle

\section{Introduction}

The realization of atomic-precision, advanced manufacturing (APAM) processes using acceptor and donor doping is an essential step towards integrated quantum circuitries\cite{Ward2020}. APAM process development, thus far, has largely focused on donor dopant-based devices in Si, and in particular, the precision placement of P atoms and their subsequent operation as qubits\cite{Weber2014}. More recently, acceptor-based Si devices have been proposed\cite{Salfi:2016, Salfi_PRL:2016, Shim:2014, Blase:2009} and realized\cite{Fuhrer:diborane} demonstrating the opportunities of hole-type devices and the need for further development of alternate chemistries for device fabrication. While APAM acceptor-doped devices have been demonstrated in Si(100) using diborane (B$_{2}$H$_{6}$)\cite{Fuhrer:diborane}, it results in the formation of electrically inactive dopant complexes due to its inherent dimerized nature\cite{Campbell2021}, prompting further investigations towards finding a more ideally suited B dopant precursor.

The use of BCl$_{3}$ as a dopant precursor has been well documented on Si substrates\cite{Herner2004, Consiglio2016, Pilli:2018, Duvauchelle2015}. Moreover, it has been used to achieve superconductivity in Si through gas immersion laser doping\cite{CAMMILLERI200875}. While these BCl$_{3}$ doping demonstrations have all been performed at temperatures or processing conditions that are inherently incompatible with the current APAM process flow, they remain suggestive of the potential viability of BCl$_{3}$ as an APAM acceptor precursor that warrants further investigation.

In this paper, we demonstrate the formation and characterization of B-doped $\delta$-layers from BCl$_{3}$-dosed Si(100). We find that BCl$_{3}$ adsorbs readily onto the Si(100) surface at room temperature. Further, the adsorption of BCl$_{3}$ generates an active carrier concentration exceeding 10$^{14}$~cm$^{-2}$ without directly annealing the substrate to facilitate activation. STM was used to verify both BCl$_{3}$ adsorption and B incorporation, while secondary ion mass spectrometry (SIMS) depth profiling was used to quantify the B content of the $\delta$-layers and verify resist selectivity. We find BCl$_{3}$ to demonstrate selectivity in adsorption on H- and Cl-terminated Si with respect to bare Si(100), making it chemically compatible with current STM lithographic processes. Hall bar measurements obtained at 3.0~K revealed similar values for sheet resistance, carrier concentration, and carrier mobility to those reported for the well developed process of P-doped $\delta$-layers formation using PH$_{3}$. This work serves as a stepping stone to enable further development of acceptor-based devices, including complementary digital electronics at the atomic-scale.

\section{Experimental Methods}

\subsection{B $\delta$-Layer Preparation}

BCl$_{3}$ dosing, STM imaging, and lithography were performed in a ScientaOmicron~VT-STM ultra-high vacuum (UHV) system with base pressure $P$ $< 2.7~\times~10^{-9}$~Pa ($2.0~\times~10^{-11}$~Torr). The STM is run using a ZyVector STM Lithography control system. The Si(100) wafers used in these experiments were p-type (B-doped) obtained from ITME with resistivity $\rho$ = 1~$\Omega\cdot$cm to 10~$\Omega\cdot$cm, and oriented within 0.5$^{\circ}$ of (100). Si samples used for dosing tests and electrical measurements were cut to 4~mm $\times$ 12~mm in size and cleaned by sonication in acetone, methanol, and isopropanol. Samples were then immediately mounted on a ScientaOmicron XA sample plate and loaded into the UHV system. Samples used for lithography tests were prepared differently. Fiducial marks were etched in the surface to allow a patterned area to be relocated in the STM (see Supplemental Information). These samples were then diced to 5~mm $\times$ 9~mm and chemically cleaned of metals and organics using a common cleaning recipe consisting of piranha solution (H$_{2}$SO$_{4}$:H$_{2}$O$_{2}$), HF, and SC-2 (H$_{2}$O:HCl:H$_{2}$O$_{2}$) before being sonicated and loaded in a similar manner to other samples. Clean Si(100)-(2$\times$1) surfaces were prepared in UHV following the procedure described in Ref.~\onlinecite{Trenhaile_O:2006} by flash annealing the sample to 1225~$^{\circ}$C. Sample temperature was monitored with an optical pyrometer. Each sample was initially scanned in the STM to check that it was free of metal and carbon contaminants then re-flashed to produce a fresh Si(100) surface prior to BCl$_{3}$ dosing.

Samples were generally slightly above room temperature during sample transfer to the dosing chamber and BCl$_{3}$ dosing to minimize water contamination\cite{Yu:2008} and, ultimately, O content within the B $\delta$-layer. BCl$_{3}$ was introduced into the UHV chamber via a precision leak valve with dosing pressures ranging from 1.3~$\times~10^{-7}$~Pa ($1.0~\times~10^{-9}$~Torr) and 5.3~$\times~10^{-6}$ Pa ($4.0~\times~10^{-8}$~Torr). Dosing values reported here are expressed in units of Langmuir (1~L = 10$^{-6}$~Torr$\cdot$s) and generally have a relative uncertainty of $\sim$0.1. Dosing parameters were varied for three samples used for electrical measurements, denoted samples B2, B3, and B4. Sample B2 was dosed with 3.6~L BCl$_{3}$ at room temperature ($\sim$20~$^{\circ}$C), sample B3 was dosed with 3.6~L BCl$_{3}$ at 350~$^{\circ}$C to try to drive chemisorption and increase the B concentration beyond the room temperature saturation, and sample B4 was dosed with 0.12~L BCl$_{3}$ at room temperature ($\sim$20~$^{\circ}$C). B incorporation anneals were not performed on these samples. Another sample used for SIMS analysis, denoted B1, was dosed with 30~L BCl$_{3}$ at 150~$^{\circ}$C to prevent any water adsorption then annealed at 350~$^{\circ}$C to incorporate the B. To form buried B $\delta$-layers and enable \textit{ex situ} SIMS and electrical measurements, BCl$_{3}$-dosed samples were encapsulated with 20~nm to 65~nm of Si at a typical rate of $\sim$0.016~nm/s using a MBE Komponenten SUSI-63 sublimation source. Samples B1 and B4 was encapsulated with a deposition rate of $\sim$0.004~nm/s at $\sim$240~$^{\circ}$C. Other samples were not heated deliberately during deposition, although thermal radiation from the deposition source heated the sample to around 200~$^{\circ}$C to 250~$^{\circ}$C. SIMS depth profiling (Eurofins EAG Materials Science) was used to quantify the B, Cl, and H content of the B $\delta$-layers and Si films.

\subsection{Electrical Measurements}

Electrical transport properties of the B $\delta$-layers were characterized with standard Hall bar measurements. Hall bars of width 25~$\mu$m and length 60~$\mu$m were patterned by electron beam lithography and the surrounding area was etched to a depth of around 100~nm with a reactive ion etcher (RIE) (see Supplemental Information). A 50 $\times$ 50 array of 1~$\mu$m sized squares were also etched into the contact pad region to enhance the contact quality between each pad and the B-doped $\delta$-layer. Ti/Au (5~nm/150~nm) was then deposited using an electron-beam deposition system to define the contact pads. Through this method, contact resistances of $<$ 1~k$\Omega$ at room temperature were routinely achieved without any annealing process. Electrical transport measurements of the B $\delta$-layers were performed at temperatures ranging from 300~K to 1.9~K and magnetic fields from -14~T to +14~T to characterize the sheet resistance ($R_{\mathrm{s}}$), hole carrier density ($n$), and mobility ($\mu$). The values for these parameters reported here were measured at 3.0~K, where the measurement system was more stable.

\section{Results and Discussion}

\subsection{BCl$_{3}$ Adsorption and Incorporation}

\begin{figure*}
\includegraphics[width=\textwidth]{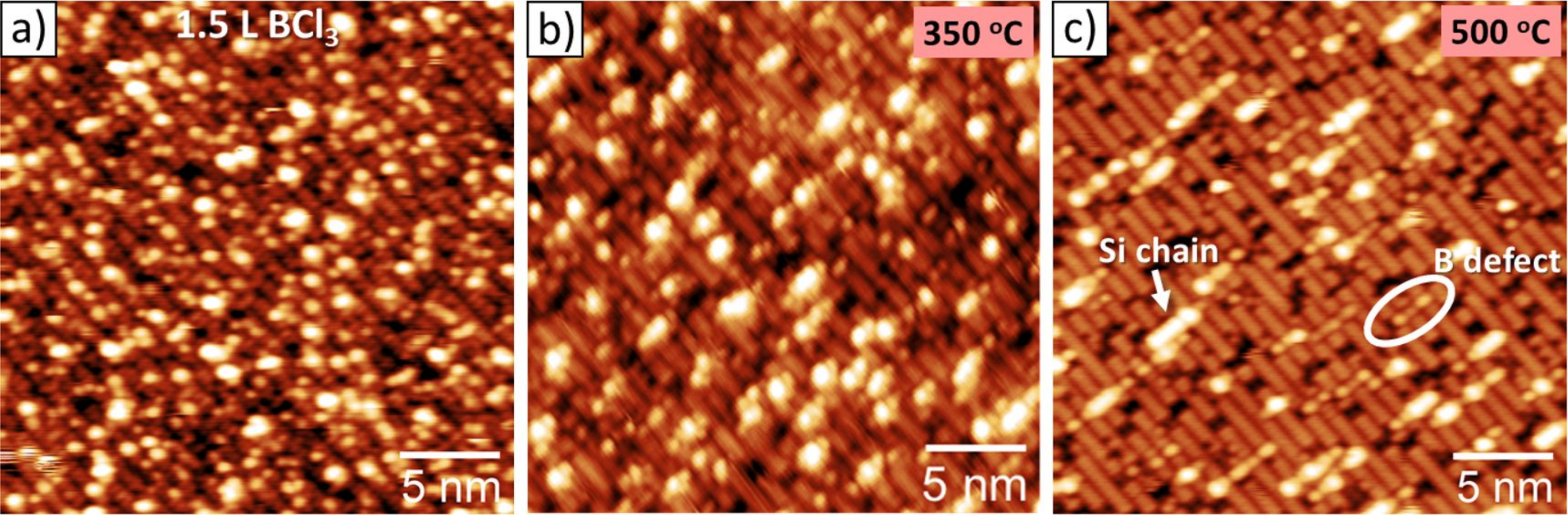}\hfill%
\caption{Filled-state STM images (a: -2.30~V, 0.15~nA, b: -2.25~V, 0.50~nA, c: -2.10~V, 0.15~nA) of a Si(100) surface dosed with 1.5~L of BCl$_{3}$ at room temperature before (a) and after annealing to 350~$^{\circ}$C (b) or 500~$^{\circ}$C (c) for 1 minute. (a)~Various adsorbed BCl$_{3}$-related species can be seen in the image as bright, round features sitting on the Si dimer rows. Many are likely chemisorbed BCl$_{2}$ and Cl atoms. These larger adsorbates amount to a surface coverage of $\sim$0.32. (b) A similarly dosed surface to (a) after annealing to 350~$^{\circ}$C showing fewer adsorbate features compared to (a). Dark dimer row defects are also seen. (c) The same surface as in (a) is shown after annealing to 500~$^{\circ}$C. Short island chains (arrow) form from ejected Si during B incorporation. Also visible are other defect features associated with subsurface B atoms (circle).}%
\label{BCl3_STM}%
\end{figure*}

Density functional theory calculations have determined that BCl$_{3}$ dissociatively chemisorbs onto Si(100) forming Si-BCl$_{2}$ and Si-Cl moieties on a single Si dimer\cite{Ferguson:2009}. There are a number of papers discussing the use of BCl$_{3}$ as a dopant source\cite{deutsch1981electrical,kunii2004situ}, and there have also been a number of investigations of adsorption on Si(111)\cite{lapiano1992chemisorption}, but to the best of our knowledge, no STM studies exist of BCl$_{3}$ adsorption on the Si(100) surface. Figure~\ref{BCl3_STM} shows filled-state STM images of a Si(100) surface exposed to BCl$_{3}$ before (a) and after annealing to 350~$^{\circ}$C (b) or 500~$^{\circ}$C (c). The surface was dosed with 1.5~L of BCl$_{3}$ at room temperature ($\sim$20~$^{\circ}$C) resulting in numerous bright features due to adsorbates as seen in Fig.~\ref{BCl3_STM}(a). BCl$_{3}$ is widely used in ALD applications and is known to stick readily to Si(100)\cite{Pilli:2018, Consiglio2016, Ferguson2002}. We also find BCl$_{3}$ to adsorb readily onto the Si(100) surface at room temperature forming a variety of surface bound structures. Many of these features are likely chemisorbed BCl$_{2}$ and as well as Cl atoms. The brighter BCl$_{3}$ features in (a) comprise a surface coverage of roughly 0.32.

After dosing, several samples were annealed at different temperatures to observe the evolution of surface features with temperature, eventually leading to incorporation of the B atoms into the substrate, which is typically done to foster electrical activation. Figure~\ref{BCl3_STM}(b) depicts the surface of a sample with the same dose as in (a) after annealing at 350~$^{\circ}$C for 1~minute. Some of the smaller round BCl$_{3}$ adsorbate features seen in (a) are gone while larger features remain. This may indicate partial dissociation or incorporation into the surface at this temperature. Also seen are darker dimer row defects, which can indicate additional Cl on the surface.

Figure~\ref{BCl3_STM}(c) depicts the surface shown in (a) after annealing at 500~$^{\circ}$C for 1~minute. This anneal was chosen based on similar anneals used for B$_{2}$H$_{6}$ \cite{Fuhrer:diborane} and AlCl$_{3}$ incorporation\cite{Radue2021}. The annealed surface appears quite differently with many of the bright round features in (a) disappearing. Short Si chains running perpendicular to the underlying dimer row direction can be seen on the surface (arrow) with their height verified to be one atomic layer. This is a potential indicator that B atoms were incorporated into the surface causing Si atoms to be ejected as adatoms, forming the chains. It is also possible that some of the ejected Si observed was the result of surface etching due to the presence of Cl on the surface during the anneal step. Very little Cl is observed on the annealed surface in Fig.~\ref{BCl3_STM}(c), matching our expectation that Cl is removed at 500~$^{\circ}$C\cite{Radue2021}. Additionally, among the various defects that are visible, there appears to be B-related surface defects (circle) which span three dimer rows and are similar to features often observed on B-doped Si substrates caused by B atoms several layers below the surface\cite{Liu2008, Wang1995}. Annealing at intermediate temperatures of 400~$^{\circ}$C and 450~$^{\circ}$C produce similar surfaces to the 350~$^{\circ}$C surface in Fig.~\ref{BCl3_STM}(b), though with additional Si chains appearing at 450~$^{\circ}$C, indicating that this temperature is closer to the onset of B incorporation.

\subsection{SIMS Characterization of B $\delta$-Layers}

\begin{figure}
\includegraphics[width=\columnwidth]{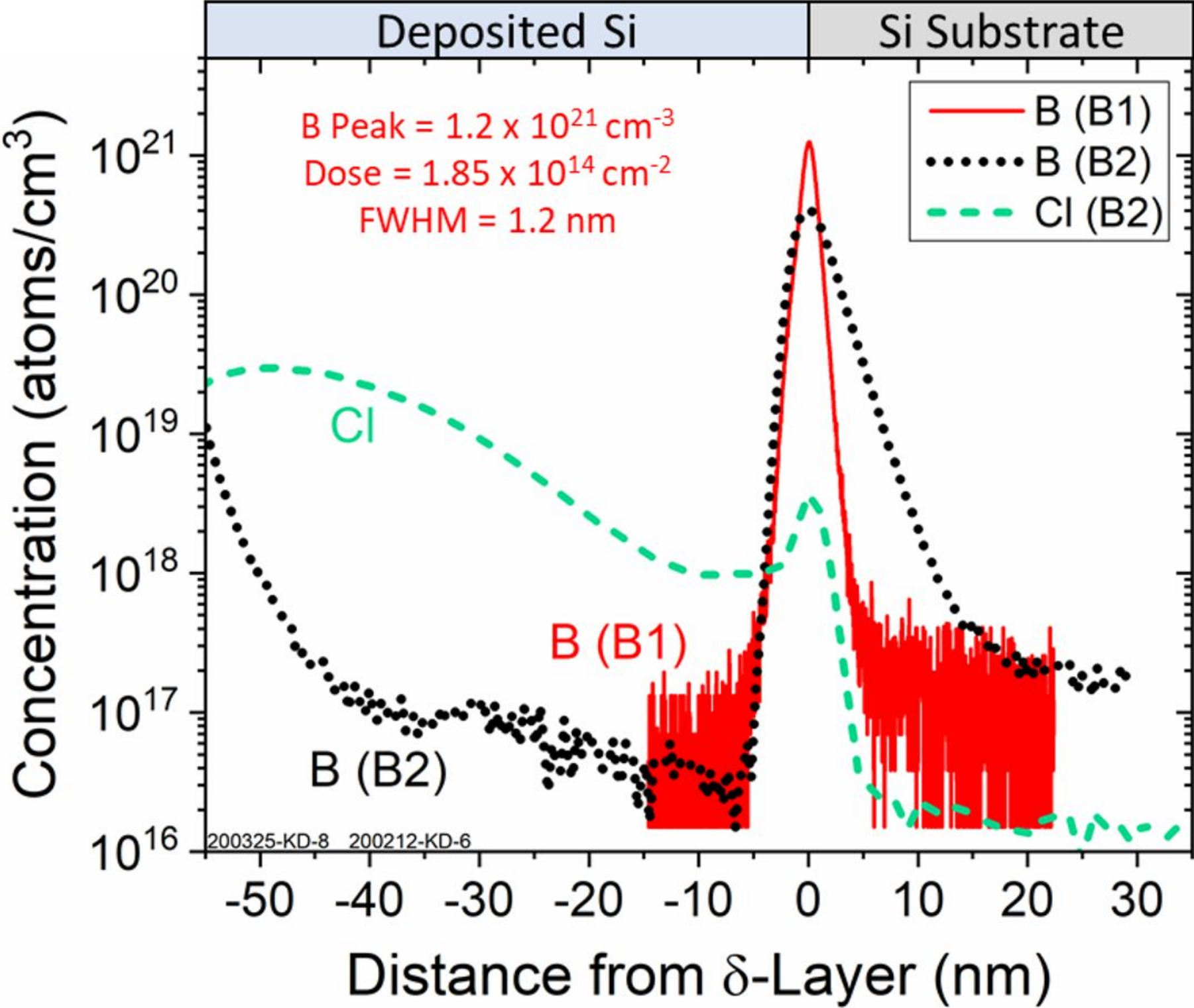}\hfill%
\caption{SIMS depth profiles of sample B2 dosed with 3.6~L BCl$_{3}$ at room temperature and sample B1 dosed with 30~L BCl$_{3}$ at 150~$^{\circ}$C. B2 was capped with $\sim$62~nm Si while B1 had a $\sim$19~nm Si cap. The B profile for B2 (dotted line) was measured under typical settings while a high resolution measurement was done for B1 (solid line). B profiles show total B ($^{10}$B + $^{11}$B). The peak B concentration for B2 is 4.1(1)~$\times$~ 10$^{20}$~cm$^{-3}$ at the interface, and for B1, the peak B concentration is 1.2(1)~$\times$~10$^{21}$~cm$^{-3}$. The B $\delta$-layer thickness is measured to be $\sim$3.4~nm for B2 based on the full width at half max of the peak and $\sim$1.2~nm for B1. A Cl depth profile from B2 (dashed line) shows some Cl present at the B layer, with the majority diffusing upwards towards the Si growth surface.}%
\label{BCl3_SIMS}%
\end{figure}

As discussed earlier, B $\delta$-doped samples were formed by exposing clean Si(100) to BCl$_{3}$ and then capping them with epitaxial Si. The $\delta$-layers were characterized \textit{ex situ} with SIMS to ascertain the estimated peak dopant concentration, the areal dose, and the estimated thickness of the $\delta$-doped layer. Figure~\ref{BCl3_SIMS} shows SIMS depth profiles of $\delta$-layer samples B1 and B2. Sample B1 was capped with $\sim$19~nm Si and B2 was capped with $\sim$62~nm Si after dosing. B profiles show in Fig.~\ref{BCl3_SIMS} represent the total B, i.e. the sum of $^{10}$B and $^{11}$B. For B2 (dotted line), the peak B concentration was measured to be 4.1(1)~$\times$~10$^{20}$~cm$^{-3}$ at the interface with a total areal dose of 1.59(1)~$\times$~10$^{14}$~cm$^{-2}$. While the SIMS measurement for B2 was performed using typical settings, $\delta$-layer sample B1 was measured using a lower primary ion energy of 250~eV resulting in a higher spacial resolution. The measurement on B1 was done as a reference to more accurately gauge the peak B concentration and, importantly, the $\delta$-layer thickness when eliminating some known SIMS measurement artifacts\cite{Gautier1996}. Areal dose accuracy, however, is consistent between different measurement types, according to the analysis provider. We believe the 3.6~L dose of B2 is at or near a saturation dose and thus the higher 30~L dose of B1 should not lead to a significantly increased B content. The high resolution measurement of B1 yielded a peak B concentration of 1.2(1)~$\times$~10$^{21}$~cm$^{-3}$ with a total areal dose of 1.85(1)~$\times$~10$^{14}$~cm$^{-2}$, similar to the areal dose of B2. Additionally, a SIMS measurement of B1 using a primary ion energy of 1~keV, similar to that of B2, shows a significant lowering of the peak to 6.3(1)~$\times$~10$^{20}$~cm$^{-3}$ (see Supplemental Information). It can therefore be assumed that the peak B concentration for sample B2 used in electrical measurements is closer to 10$^{21}$~cm$^{-3}$ than that measurement indicated. This B concentration is similar to reported concentrations for B and P $\delta$-layers deposited via a precursor of roughly 10$^{21}$~cm$^{-3}$, and the areal dose is similar for P\cite{Hagmann2018, Fuhrer:diborane}. It is also consistent with a number of reported B-doped $\delta$-layers fabricated using molecular beam epitaxy (MBE)\cite{Weir1994}.

One measure of the thickness of the $\delta$-layers is determined from the full width at half max (FWHM) of the B peak in Fig.~\ref{BCl3_SIMS}. For sample B1, the thickness of the B $\delta$-layer is estimated from the displayed depth profile to be $<$ 1.2~nm. This value is close to the measurement limit of SIMS even using high resolution settings and so may yet be an overestimate. The lower resolution measurement of sample B2 resulted in a FWHM of $\sim$3.4~nm. However, by comparing the SIMS data from multiple measurement energies, a measure of the SIMS instrument artifacts that cause broadening of the $\delta$-layer can be determined\cite{Gautier1996}. Then, an estimate of the ``true" thickness can be extrapolated for a situation with zero instrument artifact effects (see Supplemental Information). Extrapolating the measurements from sample B1 in this manner yielded a FWHM of $\sim$0.8~nm. Further, by fitting the depth profiles at each energy to a $\delta$-layer model that accounts for the instrument broadening, the standard deviation on the distribution of B can also be extrapolated to reduce the broadening effect. This type of analysis was done for sample B1 and resulted in a standard deviation of the $\delta$-layer thickness of $<$ 0.5~nm, or within 4~monolayers (ML). Previous studies of B deposited as a $\delta$-layer by MBE also found the B spreads over just a few ML\cite{Weir1994}. Taken together, these estimates give us confidence to place an upper bounds on the full thickness of our B $\delta$-layers of $<$ 1.0~nm. The thickness of the electrically active region of a $\delta$-layer can also be measured by studying the weak localization in the sample, as has been demonstrated with P $\delta$-layers\cite{Hagmann2018, Hagmann2020}, and will be explored in future studies.

\begin{figure*}[t]
\includegraphics[width=\textwidth]{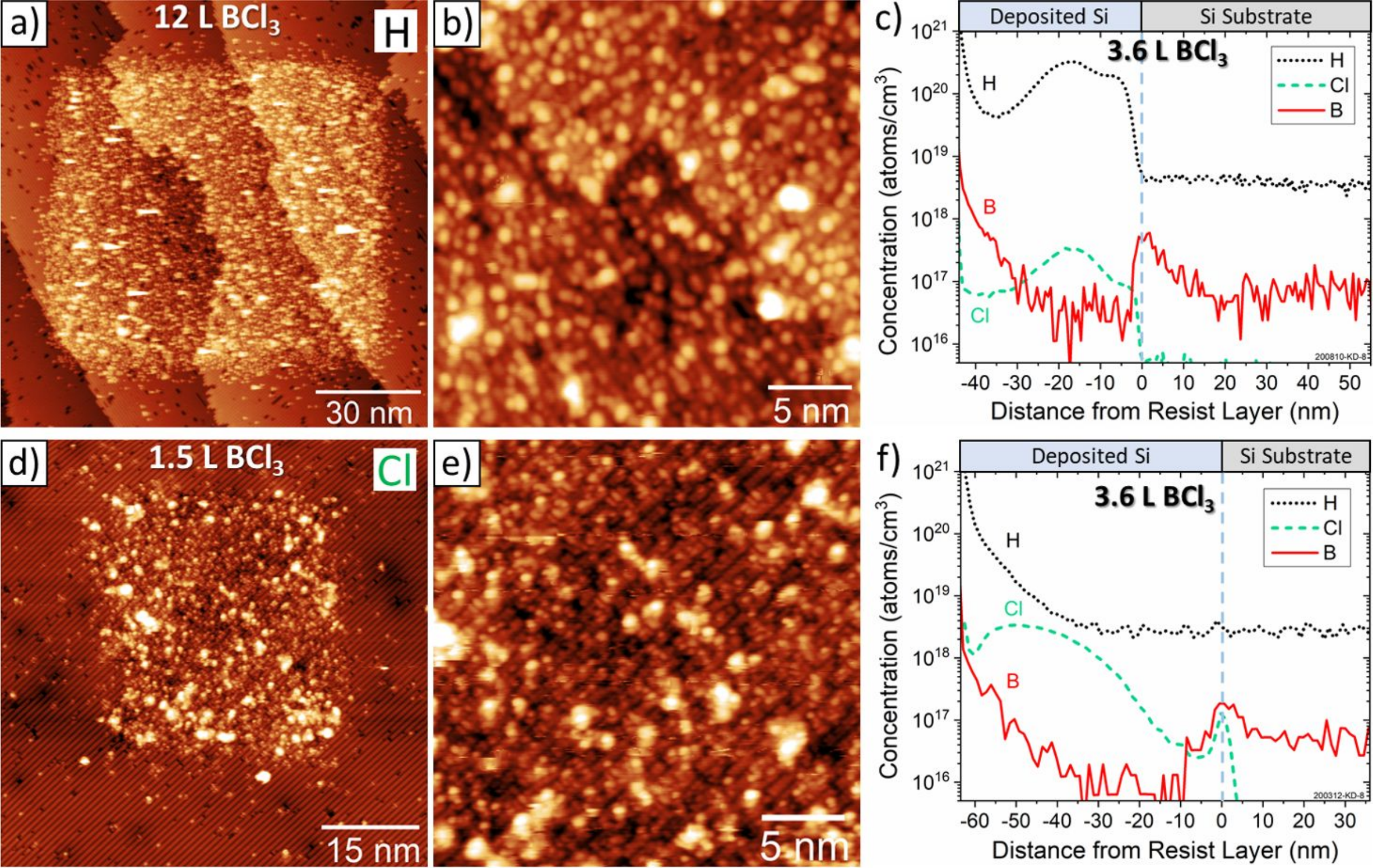}\hfill%
\caption{(a) and (b) Filled-state STM images (a: -2.22~V, 0.11~nA, b: -2.22~V, 0.13~nA) of a lithographically patterned H-Si(100) surface exposed to a 12~L BCl$_{3}$ dose. (a) Shows an 80~nm $\times$ 80~nm square filled in with BCl$_{3}$-related adsorbates, with virtually none observed outside the pattern. (b) A magnified image of the center of the dosed square in (a) showing BCl$_{3}$ adsorbates and a surface step. (c) A SIMS depth profile of a H-Si(100) sample dosed with 3.6~L of BCl$_{3}$ and capped with $\sim$45~nm of Si. BCl$_{3}$ adsorption is blocked resulting in minimal B concentration at the interface of 6.0(2)~$\times$~10$^{17}$~cm$^{-3}$. A small amount of Cl is detected in the film while the large H content indicates an intact H resist. (d) and (e) Filled-state STM images (d: -1.80~V, 0.15~nA, e: -1.80~V, 0.15~nA) of a lithographically patterned Cl-Si(100) surface exposed to a 1.5~L BCl$_{3}$ dose. (d) Shows a 25~nm $\times$ 25~nm square filled in with BCl$_{3}$-related adsorbates with virtually no adsorbates observed outside the pattern. (e) A magnified view of the image in (d) showing individual adsorbate features. (f) A SIMS depth profile of a Cl-Si(100) sample dosed with 3.6~L of BCl$_{3}$ and capped with $\sim$65~nm of Si. BCl$_{3}$ is blocked resulting in a peak B concentration of 1.8(2)~$\times$~10$^{17}$~cm$^{-3}$. The Cl content in the film results from the Cl resist diffusing upwards during Si deposition.}%
\label{B_Selectivity}%
\end{figure*}

A Cl depth profile (dashed line) from sample B2 was also measured and revealed Cl present both at the buried interface layer and throughout the Si capping layer in Fig.~\ref{BCl3_SIMS}. At the buried interface, we find an areal dose of Cl of 1.4(1)~$\times$~10$^{12}$~cm$^{-2}$, or roughly 2~$\times$~ 10$^{-3}$~ML. While there is a measured peak Cl concentration of 3.3(3)~$\times$~10$^{18}$~cm$^{-3}$ at the buried interface, a substantially larger amount of Cl was found within the epitaxial Si layer closer to the surface. Cl is expected to act as a surfactant during Si growth and exchange places with deposited Si, altering the growth kinetics\cite{pavlova2019ab} in much the same way as H, leading to the profile seen in Fig.~\ref{BCl3_SIMS}. Surfactant-mediated growth is known to enhance layer-by-layer growth in homoepitaxy and suppress three-dimensional (3D) islanding\cite{zhang1994atomic, voigtlander1994surfactant}. In addition, theoretical investigations have suggested a pathway towards removing Cl from the growth front due to lowered barriers to desorption for SiCl$_{2}$\cite{pavlova2019ab}. Under the growth conditions utilized in these investigations, we find evidence of Cl exchanging with Si adatoms and floating to the surface, but are unable to quantify amount of Cl lost. Additional optimization of the growth and incorporation parameters can significantly reduce the residual Cl at the interface as well as in the Si film. These parameters will be explored in future studies.

\subsection{Selectivity of BCl$_{3}$ with H and Cl Atomic Resists}

For compatibility with existing APAM processes, BCl$_{3}$ must demonstrate selective adsorption with respect to an atomic resist. H-Si(100) has been the standard substrate for atomically-precise device fabrication due in large part to its ease of patterning and the use of PH$_{3}$ and, more recently, B$_{2}$H$_{6}$ for doping reactions. H-based dopant precursors experience no driving force to react with the H-terminated substrate. In contrast, the use of BCl$_{3}$ presents a possible pathway for reaction with the H resist layer through abstraction\cite{Ferng2009, Li:2011} thereby compromising the selectivity required for device fabrication with good pattern fidelity.

Recent experiments demonstrating compatibility of TiCl$_{4}$ with a patterned H-resist\cite{Mitsui:1999, TiCl4:Zyvex, TiCl4:Zyvex2} at room temperature as well as density functional theory calculations predicting PH$_{3}$ compatibility with a Cl-resist\cite{Pavlova:2018} suggest BCl$_{3}$ may be similarly compatible with H-Si(100). However, this has not yet been explored experimentally prior to this work. Additionally, XPS investigations of BCl$_{3}$ adsorption on Si(100) found that Cl hindered further adsorption\cite{Pilli:2018} and atomically precise lithographic patterning of Cl-Si(100) was recently demonstrated\cite{ClLitho} providing another potential resist for B-doped APAM applications. As such, the selective adsorption of BCl$_{3}$ with both H- and Cl-based resist layers and their ability to mask B deposition on Si(100) was explored here using STM and SIMS.

Figure~\ref{B_Selectivity}(a) and (d) show filled-state STM images of lithographically patterned squares on H- and Cl-Si(100), respectively, after exposure to BCl$_{3}$ at room temperature ($\sim$20~$^{\circ}$C). The H pattern in Fig.~\ref{B_Selectivity}(a) consisted of a 80~nm $\times$ 80~nm square (lithographic parameters: +7.0~V, 1.5~nA, 4~mC/cm) that was exposed to a 12~L dose of BCl$_{3}$. While this dose is beyond our typical saturation dose, it provides a more robust test of the H pattern fidelity. The Cl pattern in (d) consisted of a 25~nm $\times$ 25~nm square (lithographic parameters: +10.0~V, 9.0~nA, 15~mC/cm) that was exposed to a more typical 1.5~L of BCl$_{3}$. The initial H- and Cl-Si(100) surfaces were first scanned with STM to verify full surface passivation. We observe that the areas outside of the patterned squares in Fig.~\ref{B_Selectivity}(a) and (d) appear as typical H- and Cl-terminated surfaces, with the exception of some spurious dangling bonds from the depassivation process. STM images obtained away from the patterned areas after BCl$_{3}$ exposure revealed that full H and Cl passivation was maintained with negligible changes in the defect concentration (see Supplemental Information). Higher magnification images of the centers of the dosed H and Cl patterns are shown in Fig.~\ref{B_Selectivity}(b) and (e), respectively. Within the patterned area, we find a variety of features similar in appearance to those seen on the Si surface in Fig.~\ref{BCl3_STM}(a). The clear preservation of the lithographic patterns after dosing verifies the ability of both H and Cl to mask the surface and validates their use as atomic resists for area-selective deposition of BCl$_{3}$. However, the STM images do not provide a quantifiable measure of the selectivity of the adsorption on the two resist surfaces.

A more precise measure of the adsorption selectivity of BCl$_{3}$ on H- and Cl-Si(100) compared to bare Si(100) was determined using SIMS. Figures~\ref{B_Selectivity}(c) and (f) present SIMS depth profiles of B, H, and Cl from BCl$_{3}$-dosed H- and Cl-Si(100) surfaces, respectively. Both fully-terminated surfaces were exposed to 3.6~L of BCl$_{3}$ at room temperature ($\sim$20~$^{\circ}$C) and encapsulated with Si to bury the interface following the same procedure used to create the B-doped $\delta$-layer samples, as discussed earlier. They were then removed from UHV and analyzed using SIMS depth profiling \textit{ex situ}. The H-Si(100) sample was not heated before encapsulation with Si, while the Cl-Si(100) sample was heated to $\sim$250~$^{\circ}$C for 10 minutes in an effort to drive off potential physisorbed species\cite{silva2020reaction}. However, both sample processing is still comparable as both were heated radiatively by the Si source during deposition to roughly this temperature. The depth profiles show a background B concentration in the substrate of $\sim$10\textsuperscript{17}~cm\textsuperscript{-3} with buried interfaces located $\sim$45~nm and $\sim$65~nm beneath the surface of the Si capping layer on the H-Si(100) and Cl-Si(100) samples, respectively. At the buried interface, a slight increase in the B concentration above 10$^{17}$~cm$^{-3}$ is observed on both samples, but it is not possible to conclude from the B signal alone whether or not this increase is due to BCl$_{3}$ adsorption or surface segregation of B from the substrate during sample preparation\cite{Zhang1996}. We have observed similar peaks in the B concentration at the Si overgrowth interface for some samples that were not exposed to BCl$_{3}$ (see Supplemental Information).

However, Cl is observed in Fig.~\ref{B_Selectivity}(c) above 10$^{17}$~cm$^{-3}$ within the Si capping layer indicating possible BCl$_{3}$ reactions on H-Si(100). To test if the increase in B was due to a reaction with H and not adsorption at single dangling bonds present on the initial H-Si surface, a similarly prepared H-terminated Si surface was exposed to the same dose of BCl$_{3}$ at $\sim$250~$^{\circ}$C. SIMS of this sample revealed slightly higher concentrations of both B and Cl, indicating a thermally activated reaction, as this dosing temperature is well below the hydrogen desorption temperature (see Supplemental Information).

Cl is also observed at the interface of the Cl-Si(100) surface and in the Si capping layer at over 10$^{18}$~cm$^{-3}$ in Fig.~\ref{B_Selectivity}(f). Unlike for the case of the H-Si(100) surface, it cannot be assumed that the presence this Cl is due to reactions with BCl$_{3}$ as the surface was already saturated with Cl. We cannot determine quantitatively how much, if any, BCl$_{3}$ reacts through the Cl resist other than it appears negligible and is several orders of magnitude lower than that seen on bare Si(100) as in Fig.~\ref{BCl3_SIMS}. As Fig.~\ref{B_Selectivity}(c) and (f) show, Cl readily migrates from the interface towards the growth surface during Si overgrowth. Even for the case of Cl-Si(100), with an initial Cl coverage approximately 1~ML, the buried interface after Si growth is left with and areal coverage of just 5.1(2)~$\times$~10$^{10}$~cm$^{-2}$ of Cl, or $\sim$$10^{-4}$~ML. While the residual Cl content at the buried interface is not insignificant, i.e. it is above the background level in the substrate, the magneto-transport and device measurements presented in the following section suggests Cl does not have a detrimental impact on Hall bar measurements.

To more accurately determine the adsorption selectivity of H and Cl resists with BCl$_{3}$, we compare the total areal dose of B measured by SIMS for both resists to that of bare Si(100) when all three are dosed with 3.6~L BCl$_{3}$. The SIMS measurement in Fig.~\ref{BCl3_SIMS} (lower resolution B2 measurement) gives the bare Si B areal dose of 1.59(1)~$\times$~10\textsuperscript{14}~cm\textsuperscript{-2}, while the B areal doses on the resist layers are measured from Fig.~\ref{B_Selectivity}(c) and (f) to be 2.6(1)~$\times$~10\textsuperscript{11}~cm\textsuperscript{-2} for H-Si(100) and 5.2(2)~$\times$~10\textsuperscript{10}~cm\textsuperscript{-2} for Cl-Si(100). The adsorption selectivity ratios for these resists is then calculated in a similar manner as is done for area-selective deposition\cite{Mackus2019, Gladfelter1993}:
\begin{equation}
\mathrm{Selectivity~Ratio} = \left[1 - \left(\frac{\theta_{\mathrm{GA}}-\theta_{\mathrm{NGA}}}{\theta_{\mathrm{GA}}+\theta_{\mathrm{NGA}}}\right)\right]^{-1},
\label{Eq:selectivity}
\end{equation}
where $\theta_{\mathrm{GA}}$ is taken to be the measured B dose in the growth area (bare Si) and $\theta_{\mathrm{NGA}}$ is the B dose in the non-growth area (H- and Cl-terminated Si). From Eq.~(\ref{Eq:selectivity}), the adsorption selectivity is found to be 310(10):1 and 1529(5):1 for H- and Cl-Si(100) at room temperature, respectively.

\begin{figure*}[th]
\includegraphics[width=\textwidth]{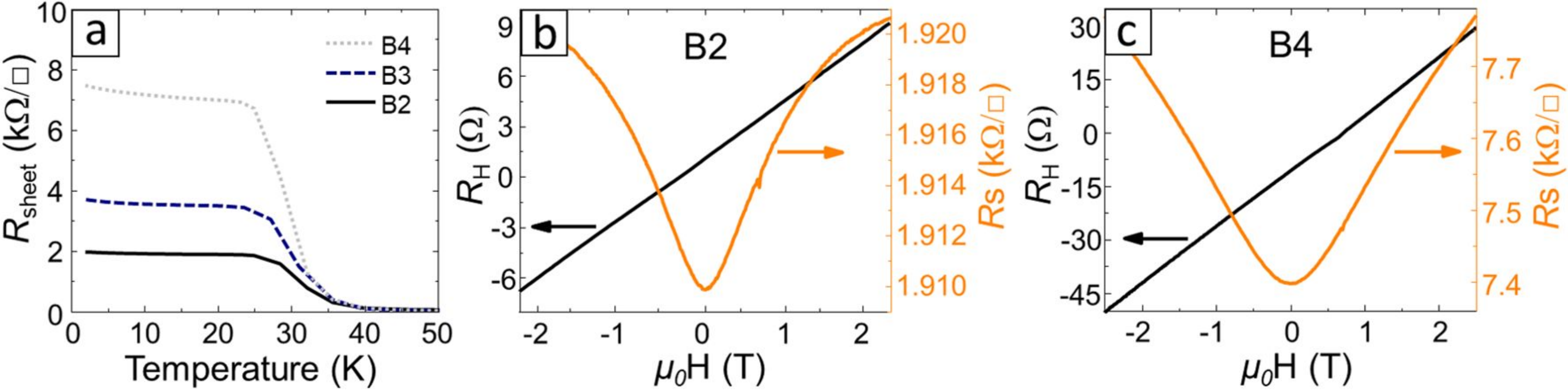}\hfill%
\caption{Electric transport measurements of B-doped $\delta$-layers fabricated into 25~$\mu$m $\times$ 60~$\mu$m mesa etched Hall bars. (a) The temperature dependence of sheet resistance between \mbox{50~K} and 1.9~K is shown for B2 (solid line), B3 (dashed line) and B4 (dotted line). Comparing dose temperature, sample B2 (room temperature dose) has lower $R_{\mathrm{sheet}}$ (higher conductivity) compared to sample B3 (350~$^{\circ}$C dose). Comparing dose amount, sample B4 (0.12~L dose) has a higher $R_{\mathrm{sheet}}$ (lower conductivity) than either B2 or B3 (3.6~L doses). (b) and (c) Magnetic field dependence of the longitudinal, $R_{\mathrm{s}}$ (right axis), and transverse, $R_{\mathrm{H}}$ (left axis), resistance at 3.0~K for samples B2 and B4, respectively.}%
\label{E_measurements}%
\end{figure*}

\begin{table*}
\begin{tabular}{cclcccccc}
\toprule[0.12em]
\\[-0.9em]
\textbf{Sample} && \textbf{BCl$_{3}$ Dose} && \textbf{$R_{\mathrm{s}}$ (k$\Omega\square^{-1}$)} && \textbf{$n$ (10$^{14}$~cm$^{-2}$)} && \textbf{$\mu$ (cm$^{2}$V$^{-1}$s$^{-1}$)} \\[-0.9em]
\\\toprule[0.1em]
\\[-0.9em]
B4 && 0.12 L, 20~$^{\circ}$C && 7.04 && 0.395 && 21.4 \\\hline
B3 && 3.6 L, 350~$^{\circ}$C && 3.69 && 1.13  && 15.0 \\\hline
B2 && 3.6 L, 20~$^{\circ}$C  && 1.91 && 1.90  && 38.0 \\\bottomrule[0.12em]
\end{tabular}
\caption{Electrical properties of B-doped $\delta$-layer samples measured at 3.0~K. The BCl$_{3}$ dose parameters for samples B2, B3, and B4 are shown along with the measured sheet resistance ($R_{\mathrm{s}}$), hole carrier density ($n$), and mobility ($\mu$).}
\label{Tab:Sample}
\end{table*}

A selectivity ratio of 1000:1 would be similar to values obtained for PH$_{3}$ on H-Si(100) and, as such, is the desired goal. However, a final dopant concentration due to incorporation through the resist of $<$ 10$^{18}$~cm$^{-3}$ is sufficient to ensure carrier freeze out of stray adsorbed BCl$_{3}$ within the resist layer at typical device measurement temperatures of ${\leq}$ 4~K. Both H and Cl resist measurements presented here result in B concentrations below that threshold, as seen in Fig.~\ref{B_Selectivity}(c) and (f). Still, a small amount of BCl$_{3}$ incorporation through the H resist could pose an issue for fabricating single-atom devices, though that remains to be determined for the amount measured here.

\subsection{Electrical Characterization of B $\delta$-Layers}

We studied the electrical transport characteristics of B-doped $\delta$-layer samples B2, B3, and B4, which were fabricated into mesa-etched Hall bars 25~$\mu$m $\times$ 60~$\mu$m in size, as described earlier. An optical micrograph of a Hall bar device is shown in the Supplemental Information. While full B incorporation was demonstrated by annealing at 500~$^{\circ}$C, as shown in Fig.~\ref{BCl3_STM}(c), $\delta$-layers discussed here were not annealed that hot, as described earlier. Incorporation of dopants through annealing is typically considered important for good electrical activation, but here we wanted to isolate the effects of annealing on electrical properties and actually found that B dopant activation was high despite no deliberate incorporation. The dose parameters and measurement results of sheet resistance ($R_{\mathrm{s}}$), hole carrier density ($n$), and mobility ($\mu$) for these samples are described in Table~\ref{Tab:Sample} with reported values measured at 3.0~K.

The temperature dependence of the sheet resistance between 50~K and 1.9~K is shown in Fig.~\ref{E_measurements}(a) for samples B2 (solid line), B3 (dashed line), and B4 (dotted line). The measured sheet resistance values of $R_{\mathrm{sheet}} <$~10~k$\Omega\square^{-1}$ below 30~K reflect the properties of the $\delta$-layer as the resistance of the substrate and the Si capping layer increases to over 10~M$\Omega\square^{-1}$ at these temperatures and thus do not contribute to the measurement. Figure~\ref{E_measurements}(b) and (c) show the magnetic field dependence of the longitudinal magnetoresistance, $R_{\mathrm{s}}$ (left axis), and transverse Hall resistance, $R_{\mathrm{H}}$ (right axis), resistance of samples B2 and B4. The positive slope of $R_{\mathrm{H}}$ observed in (b) and (c) indicates that holes are the primary carriers in these devices, verifying that an acceptor-doped region is being measured. Likewise, the positive values of $R_{\mathrm{s}}$ measured for these samples also indicates acceptor dopants\cite{doi:10.1063/1.5045338}, as expected for a B $\delta$-layer. Since APAM processes requires tight control over the entire doping and annealing process, achieving a minimum sheet resistance for sample B2 of $R_{\mathrm{sheet}}$ = 1.91~k$\Omega\square^{-1}$ without an incorporation annealing step validates the use of BCl$_{3}$ as an acceptor doping precursor. Comparing this result to $\delta$-layer devices dosed with PH$_{3}$ shows they are electrically similar with the sheet resistances of P-doped $\delta$-layers reported to be between $R_{\mathrm{sheet}}$ = 0.5~k$\Omega\square^{-1}$ and $R_{\mathrm{sheet}}$ = 5~k$\Omega\square^{-1}$ for mildly dosed sample\cite{goh2006influence, weber2012ohm}.

From the data in Fig.~\ref{E_measurements}(b), the carrier density calculated for sample B2 is $n$ = 1.90~$\times$~10$^{14}$~cm$^{-2}$ with a hole mobility $\mu$ = 38.0~cm$^{2}$V$^{-1}$s$^{-1}$. This carrier density is slightly larger than the B areal dose measured by SIMS of 1.59(1)~$\times$~10$^{14}$~cm$^{-2}$, implying full dopant activation and possibly some additional electrical pathways such as vacancies or impurities in the overgrowth layer. We compare sample B2 with B4, which was produced with a smaller BCl$_{3}$ dose. The sheet resistance of B4 was larger at $R_{\mathrm{sheet}}$ = 7.04~k$\Omega\square^{-1}$ compared to B2. The carrier density and mobility for sample B4 were both lower compared to B2 with $n$ = 3.95~$\times$~10$^{13}$~cm$^{-2}$ and $\mu$ = 21.4~cm$^{2}$V$^{-1}$s$^{-1}$. The increased carrier density of B2 is only a factor of $\sim$5 times larger than for B4 despite the BCl$_{3}$ dose being a factor of 30 times larger. Suggesting that the surface of B2 may have been fully saturated by the 3.6~L dose.

Sample B3 was dosed with the same dose of BCl$_{3}$ as B2 but at an elevated temperature of 350~$^{\circ}$C to see if BCl$_{3}$ chemisorption could be facilitated in order to increase B concentration in the $\delta$-layer. As shown in Fig.~\ref{BCl3_STM}(b), BCl$_{3}$ dissociation is enhanced at this temperature relative to room temperature. However, the sheet resistance of B3 was observed to increase to $R_{\mathrm{s}}$ = 3.69~k$\Omega\square^{-1}$ compared to the room temperature-dosed ($\sim$20~$^{\circ}$C) sample B2 as shown in Fig.~\ref{E_measurements}(a), while the hole carrier density and mobility for this sample was found to decrease at $n$ = 1.13~$\times$~10$^{14}$~cm$^{-2}$ and $\mu$ = 15.0~cm$^{2}$V$^{-1}$s$^{-1}$. As with B2, a SIMS depth profile of B for B3 (not shown) measured a B areal dose similar to the carrier density of 1.17(1)~$\times$~ 10$^{14}$~cm$^{-2}$. This indicates BCl$_{3}$ adsorbs readily at room temperature and more so than at 350~$^{\circ}$C. It is possible that at 350~$^{\circ}$C, the sticking coefficient of BCl$_{3}$ is decreased resulting in fewer molecules chemisorbing. Alternatively, pairs of B atoms may be starting to dimerize as they diffuse about and incorporate into the surface at this temperature, becoming electrically inactive. It has been reported in literature that P-doped $\delta$-layers fabricated by dosing Si(100) with PH$_{3}$ at 550~$^{\circ}$C showed a similar increase in sheet resistance and decrease in carrier concentration when compared to devices dosed at room temperature\cite{goh2006influence}. Comparing the SIMS depth profile measurements of the B areal doses in samples B2 and B3 indicates that the total amount of B was reduced by a factor of 0.26. This reduction does not fully explain the drop in carrier density in B3 by a factor of 0.41 compared to B2, indicating some B deactivation may also be occurring.

Sample B2 with its more optimized dosing conditions has similar properties to some P-doped $\delta$-layers with reported Hall carrier densities of $n <$ 2~$\times$~10$^{14}$~cm$^{-2}$ and mobilities between $\mu$ = 30~cm$^{2}$V$^{-1}$s$^{-1}$ and $\mu$ = 60~cm$^{2}$V$^{-1}$s$^{-1}$\cite{goh2006influence, weber2012ohm}. Further, compared to literature values for B-doped $\delta$-layer properties\cite{Fuhrer:diborane} dosed with 1~L of B$_{2}$H$_{6}$ ($n$ $\approx 1~\times$~10$^{14}$~cm$^{-2}$, $\mu \approx$ 20~cm$^{2}$V$^{-1}$s$^{-1}$) and 170~L of B$_{2}$H$_{6}$ ($R_{\mathrm{sheet}} \approx$ 12~k$\Omega\square^{-1}$) then annealed to 250~$^{\circ}$C, these BCl$_{3}$-dosed samples show lower values of sheet resistance with B2 also having a higher carrier density and mobility without utilizing an incorporation anneal. This result supports the notion that due to B$_{2}$H$_{6}$ being deposited as B dimers, many are electrically inactive without high annealing temperatures, while deposition of single B via BCl$_{3}$ results in higher electrical activation.

\begin{figure}
\includegraphics[width=\columnwidth]{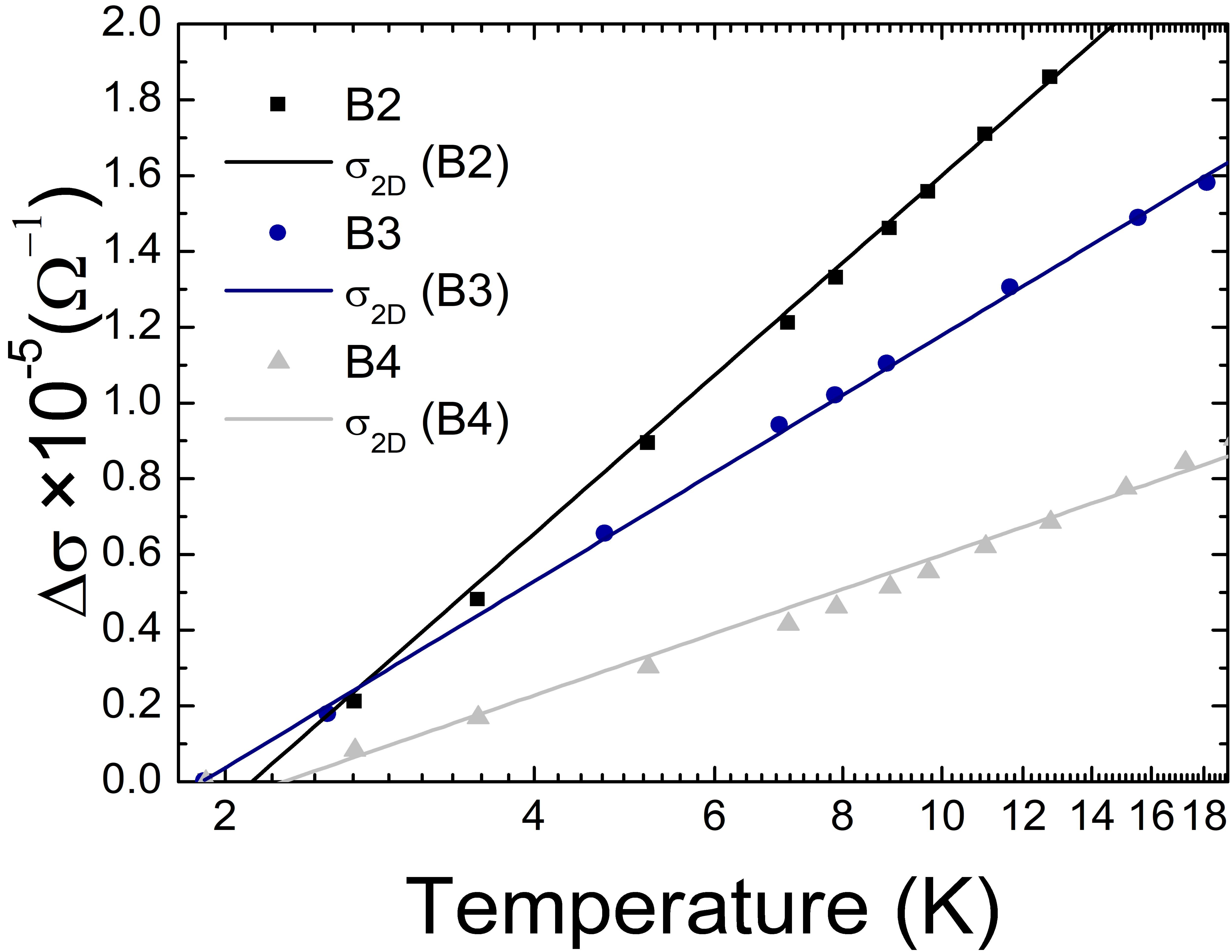}\hfill%
\caption{Temperature dependence of the relative conductivity for samples B2 (squares), B3 (circles), and B4 (triangles). Temperature is plotted on a log scale. The data for all three samples appears linear indicating a 2D nature to the conduction. A 2D conductance model (Eq.~(\ref{Eq:2D_cond})) was fit to the data to extract the values of parameter $p$. The fits yield $p$ = 0.84 (B2), $p$ = 0.56 (B3), and $p$ = 0.33 (B4).}%
\label{E_logT}%
\end{figure}

While these B-doped $\delta$-layers have favorable electrical properties, their two-dimensional (2D) nature is also critical for use in APAM and quantum information applications and requires investigation. The estimated $\delta$-layer width from the SIMS depth profile of Fig.~\ref{BCl3_SIMS} is not necessarily reliable evidence for a true 2D layer due to limitations of the SIMS measurement. Therefore, we use the electrical measurements to ascertain the dimensionality of the hole transport through the devices. We define a relative conductivity, $\Delta\sigma$, as the change in conductivity from minimum measurement temperature for each data set. This allows the data from multiple samples to be plotted together for clarity. Figure~\ref{E_logT} shows the temperature dependence of the relative conductivity in the absence of a magnetic field with temperature plotted on a log scale for samples B2 (squares), B3 (circles), and B4 (triangles).

We then fit the $\Delta\sigma(T)$ data as a function of $\ln(T)$ to a 2D conductance model\cite{lee1985disordered}.
\begin{equation}
\sigma_{\mathrm{2D}}(T) = \sigma_{0}+\frac{pe^{2}}{2\hbar\pi^{2}} \ln{\frac{T}{T_{0}}}
\label{Eq:2D_cond}
\end{equation}
The dimensionless fitting parameter $p$ depends on the scattering mechanism of the carriers in the device and the dephasing length. The fits to the data are shown as solid lines in Fig.~\ref{E_logT} with fit values for $p$ of $p$ = 0.84 for B2, $p$ = 0.56 for B3, and $p$ = 0.33 for B4. Similar values of $p$ have been reported for other B $\delta$-layers highly doped via MBE\cite{Weir1994}. Based on these fits, $\Delta$$\sigma$ is found to have a linear relationship to $\ln(T)$, as seen in Fig.~\ref{E_logT}. This indicates that the transport data is consistent with 2D conduction\cite{lee1985disordered}, however we cannot rule out other mechanisms that contribute to a linear $\ln(T)$ dependence such as electron-electron interactions based on this data alone. However, combined with the relatively thin B layer determined from the SIMS data, this analysis leads us to conclude that B-doped $\delta$-layers formed from BCl$_{3}$ dosing described here likely produces a 2D hole system suitable for fabrication of quantum devices in APAM applications. Obtaining a more accurate measure of the dimensionality of conduction requires further investigations such as weak localization measurement techniques, as have been done for P-doped $\delta$-layers\cite{Hagmann2018, Hagmann2020}.

\section{Conclusion}

In summary, we have demonstrated the viability of BCl$_{3}$ as a molecular precursor for acceptor-doping of B $\delta$-layers in Si(100) for the fabrication of atomically-precise devices. B concentrations achieved in the $\delta$-layers were shown to be similar to those achieved in the more well-developed processes for P-doped $\delta$-layers. Importantly, BCl$_{3}$ was shown to be compatible with both H- and Cl-based resists for STM lithography-based device fabrication. Cl content is not expected to be detrimental to device operation as it moves out of the buried interface towards the top of the Si epitaxial layer during growth, which can additionally be etched/polished back. Moreover, the electrical properties demonstrated here on B-doped $\delta$-layers rival those of P-doping from PH$_{3}$ and B-doping from B$_{2}$H$_{6}$ without yet optimizing the incorporation anneal step. The compatibility of BCl$_{3}$ dosing with multiple resists ensures that it can be implemented in process flows including PH$_{3}$ for the formation of exotic atomic-scale devices like pn junctions. The fact that BCl$_{3}$ shows a slight preference for reacting with H-Si over Cl-Si could be exploited for other process applications involving halogen adsorption into patterned areas on H-Si(100)\cite{Ferng2009} to serve as a negative resist. Overall, these results point towards the suitability of BCl$_{3}$ as  a viable gaseous precursor for acceptor doping processes associated with atomic precision advanced manufacturing of Si-based devices.

\section{Acknowledgement}

The authors would like to thank J. Foglebach, D. Ketchum, P. Hannah, J. Moody, S. Brown and T. Olver for their assistance with BCl$_{3}$ handling and installation. We also acknowledge Matt Radue, George Wang, Ezra Bussmann, and Scott Schmucker for thought provoking discussions. This work was supported in part by the Laboratory Directed Research and Development program at Sandia National Laboratories, a multimission laboratory managed and operated by National Technology and Engineering Solutions of Sandia, LLC., a wholly owned subsidiary of Honeywell International, Inc., for the U.S. Department of Energy's National Nuclear Security Administration under contract DE-NA-0003525. This paper describes objective technical results and analysis. Any subjective views or opinions that might be expressed in the paper do not necessarily represent the views of the U.S. Department of Energy or the United States Government.

\bibliographystyle{apsrev4-1}
\bibliography{BCl3_incorporation}

\begin{thebibliography}{45}%
\makeatletter
\providecommand \@ifxundefined [1]{%
 \@ifx{#1\undefined}
}%
\providecommand \@ifnum [1]{%
 \ifnum #1\expandafter \@firstoftwo
 \else \expandafter \@secondoftwo
 \fi
}%
\providecommand \@ifx [1]{%
 \ifx #1\expandafter \@firstoftwo
 \else \expandafter \@secondoftwo
 \fi
}%
\providecommand \natexlab [1]{#1}%
\providecommand \enquote  [1]{``#1''}%
\providecommand \bibnamefont  [1]{#1}%
\providecommand \bibfnamefont [1]{#1}%
\providecommand \citenamefont [1]{#1}%
\providecommand \href@noop [0]{\@secondoftwo}%
\providecommand \href [0]{\begingroup \@sanitize@url \@href}%
\providecommand \@href[1]{\@@startlink{#1}\@@href}%
\providecommand \@@href[1]{\endgroup#1\@@endlink}%
\providecommand \@sanitize@url [0]{\catcode `\\12\catcode `\$12\catcode
  `\&12\catcode `\#12\catcode `\^12\catcode `\_12\catcode `\%12\relax}%
\providecommand \@@startlink[1]{}%
\providecommand \@@endlink[0]{}%
\providecommand \url  [0]{\begingroup\@sanitize@url \@url }%
\providecommand \@url [1]{\endgroup\@href {#1}{\urlprefix }}%
\providecommand \urlprefix  [0]{URL }%
\providecommand \Eprint [0]{\href }%
\providecommand \doibase [0]{http://dx.doi.org/}%
\providecommand \selectlanguage [0]{\@gobble}%
\providecommand \bibinfo  [0]{\@secondoftwo}%
\providecommand \bibfield  [0]{\@secondoftwo}%
\providecommand \translation [1]{[#1]}%
\providecommand \BibitemOpen [0]{}%
\providecommand \bibitemStop [0]{}%
\providecommand \bibitemNoStop [0]{.\EOS\space}%
\providecommand \EOS [0]{\spacefactor3000\relax}%
\providecommand \BibitemShut  [1]{\csname bibitem#1\endcsname}%
\let\auto@bib@innerbib\@empty
\bibitem [{\citenamefont {Ward}\ \emph {et~al.}(2020)\citenamefont {Ward},
  \citenamefont {Schmucker}, \citenamefont {Anderson}, \citenamefont
  {Bussmann}, \citenamefont {Tracy}, \citenamefont {Lu}, \citenamefont
  {Maurer}, \citenamefont {Baczewski}, \citenamefont {Campbell}, \citenamefont
  {Marshall},\ and\ \citenamefont {Misra}}]{Ward2020}%
  \BibitemOpen
  \bibfield  {author} {\bibinfo {author} {\bibfnamefont {D.~R.}\ \bibnamefont
  {Ward}}, \bibinfo {author} {\bibfnamefont {S.~W.}\ \bibnamefont {Schmucker}},
  \bibinfo {author} {\bibfnamefont {E.~M.}\ \bibnamefont {Anderson}}, \bibinfo
  {author} {\bibfnamefont {E.}~\bibnamefont {Bussmann}}, \bibinfo {author}
  {\bibfnamefont {L.}~\bibnamefont {Tracy}}, \bibinfo {author} {\bibfnamefont
  {T.-M.}\ \bibnamefont {Lu}}, \bibinfo {author} {\bibfnamefont {L.~N.}\
  \bibnamefont {Maurer}}, \bibinfo {author} {\bibfnamefont {A.}~\bibnamefont
  {Baczewski}}, \bibinfo {author} {\bibfnamefont {D.~M.}\ \bibnamefont
  {Campbell}}, \bibinfo {author} {\bibfnamefont {M.~T.}\ \bibnamefont
  {Marshall}}, \ and\ \bibinfo {author} {\bibfnamefont {S.}~\bibnamefont
  {Misra}},\ }\href {http://arxiv.org/abs/2002.11003} {\  (\bibinfo {year}
  {2020})},\ \Eprint {http://arxiv.org/abs/2002.11003} {arXiv:2002.11003}
  \BibitemShut {NoStop}%
\bibitem [{\citenamefont {Weber}\ \emph {et~al.}(2014)\citenamefont {Weber},
  \citenamefont {Tan}, \citenamefont {Mahapatra}, \citenamefont {Watson},
  \citenamefont {Ryu}, \citenamefont {Rahman}, \citenamefont {Hollenberg},
  \citenamefont {Klimeck},\ and\ \citenamefont {Simmons}}]{Weber2014}%
  \BibitemOpen
  \bibfield  {author} {\bibinfo {author} {\bibfnamefont {B.}~\bibnamefont
  {Weber}}, \bibinfo {author} {\bibfnamefont {Y.~H.}\ \bibnamefont {Tan}},
  \bibinfo {author} {\bibfnamefont {S.}~\bibnamefont {Mahapatra}}, \bibinfo
  {author} {\bibfnamefont {T.~F.}\ \bibnamefont {Watson}}, \bibinfo {author}
  {\bibfnamefont {H.}~\bibnamefont {Ryu}}, \bibinfo {author} {\bibfnamefont
  {R.}~\bibnamefont {Rahman}}, \bibinfo {author} {\bibfnamefont {L.~C.}\
  \bibnamefont {Hollenberg}}, \bibinfo {author} {\bibfnamefont
  {G.}~\bibnamefont {Klimeck}}, \ and\ \bibinfo {author} {\bibfnamefont
  {M.~Y.}\ \bibnamefont {Simmons}},\ }\href@noop {} {\bibfield  {journal}
  {\bibinfo  {journal} {Nature Nanotechnology}\ }\textbf {\bibinfo {volume}
  {9}},\ \bibinfo {pages} {430} (\bibinfo {year} {2014})}\BibitemShut {NoStop}%
\bibitem [{\citenamefont {Salfi}\ \emph {et~al.}(2014)\citenamefont {Salfi},
  \citenamefont {Tong}, \citenamefont {Rogge},\ and\ \citenamefont
  {Culcer}}]{Salfi:2016}%
  \BibitemOpen
  \bibfield  {author} {\bibinfo {author} {\bibfnamefont {J.}~\bibnamefont
  {Salfi}}, \bibinfo {author} {\bibfnamefont {M.}~\bibnamefont {Tong}},
  \bibinfo {author} {\bibfnamefont {S.}~\bibnamefont {Rogge}}, \ and\ \bibinfo
  {author} {\bibfnamefont {D.}~\bibnamefont {Culcer}},\ }\href {\doibase
  10.1088/0957-4484/27/24/244001} {\bibfield  {journal} {\bibinfo  {journal}
  {Nanotechnology}\ }\textbf {\bibinfo {volume} {27}},\ \bibinfo {pages}
  {244001} (\bibinfo {year} {2014})}\BibitemShut {NoStop}%
\bibitem [{\citenamefont {Salfi}\ \emph {et~al.}(2016)\citenamefont {Salfi},
  \citenamefont {Mol}, \citenamefont {Culcer},\ and\ \citenamefont
  {Rogge}}]{Salfi_PRL:2016}%
  \BibitemOpen
  \bibfield  {author} {\bibinfo {author} {\bibfnamefont {J.}~\bibnamefont
  {Salfi}}, \bibinfo {author} {\bibfnamefont {J.~A.}\ \bibnamefont {Mol}},
  \bibinfo {author} {\bibfnamefont {D.}~\bibnamefont {Culcer}}, \ and\ \bibinfo
  {author} {\bibfnamefont {S.}~\bibnamefont {Rogge}},\ }\href {\doibase
  10.1103/PhysRevLett.116.246801} {\bibfield  {journal} {\bibinfo  {journal}
  {Phys. Rev. Lett.}\ }\textbf {\bibinfo {volume} {116}},\ \bibinfo {pages}
  {246801} (\bibinfo {year} {2016})}\BibitemShut {NoStop}%
\bibitem [{\citenamefont {Shim}\ and\ \citenamefont {Tahan}(2014)}]{Shim:2014}%
  \BibitemOpen
  \bibfield  {author} {\bibinfo {author} {\bibfnamefont {Y.-P.}\ \bibnamefont
  {Shim}}\ and\ \bibinfo {author} {\bibfnamefont {C.}~\bibnamefont {Tahan}},\
  }\href {\doibase 10.1038/ncomms5225} {\bibfield  {journal} {\bibinfo
  {journal} {Nat. Commun.}\ }\textbf {\bibinfo {volume} {5}},\ \bibinfo {pages}
  {4225} (\bibinfo {year} {2014})}\BibitemShut {NoStop}%
\bibitem [{\citenamefont {Blase}\ \emph {et~al.}(2009)\citenamefont {Blase},
  \citenamefont {Bustarret}, \citenamefont {Chapelier}, \citenamefont {Klein},\
  and\ \citenamefont {Marcenat}}]{Blase:2009}%
  \BibitemOpen
  \bibfield  {author} {\bibinfo {author} {\bibfnamefont {X.}~\bibnamefont
  {Blase}}, \bibinfo {author} {\bibfnamefont {E.}~\bibnamefont {Bustarret}},
  \bibinfo {author} {\bibfnamefont {C.}~\bibnamefont {Chapelier}}, \bibinfo
  {author} {\bibfnamefont {T.}~\bibnamefont {Klein}}, \ and\ \bibinfo {author}
  {\bibfnamefont {C.}~\bibnamefont {Marcenat}},\ }\href {\doibase
  10.1038/nmat2425} {\bibfield  {journal} {\bibinfo  {journal} {Nat. Mater.}\
  }\textbf {\bibinfo {volume} {8}},\ \bibinfo {pages} {375} (\bibinfo {year}
  {2009})}\BibitemShut {NoStop}%
\bibitem [{\citenamefont {Škereň}\ \emph {et~al.}(2020)\citenamefont
  {Škereň}, \citenamefont {Köster}, \citenamefont {Douhard}, \citenamefont
  {Fleischmann},\ and\ \citenamefont {Fuhrer}}]{Fuhrer:diborane}%
  \BibitemOpen
  \bibfield  {author} {\bibinfo {author} {\bibfnamefont {T.}~\bibnamefont
  {Škereň}}, \bibinfo {author} {\bibfnamefont {S.~A.}\ \bibnamefont
  {Köster}}, \bibinfo {author} {\bibfnamefont {B.}~\bibnamefont {Douhard}},
  \bibinfo {author} {\bibfnamefont {C.}~\bibnamefont {Fleischmann}}, \ and\
  \bibinfo {author} {\bibfnamefont {A.}~\bibnamefont {Fuhrer}},\ }\href
  {\doibase 10.1038/s41928-020-0445-5} {\bibfield  {journal} {\bibinfo
  {journal} {Nature Electronics}\ ,\ \bibinfo {pages} {2520}} (\bibinfo {year}
  {2020})}\BibitemShut {NoStop}%
\bibitem [{\citenamefont {Campbell}\ \emph {et~al.}(2021)\citenamefont
  {Campbell}, \citenamefont {Ivie}, \citenamefont {Bussmann}, \citenamefont
  {Schmucker}, \citenamefont {Baczewski},\ and\ \citenamefont
  {Misra}}]{Campbell2021}%
  \BibitemOpen
  \bibfield  {author} {\bibinfo {author} {\bibfnamefont {Q.}~\bibnamefont
  {Campbell}}, \bibinfo {author} {\bibfnamefont {J.~A.}\ \bibnamefont {Ivie}},
  \bibinfo {author} {\bibfnamefont {E.}~\bibnamefont {Bussmann}}, \bibinfo
  {author} {\bibfnamefont {S.~W.}\ \bibnamefont {Schmucker}}, \bibinfo {author}
  {\bibfnamefont {A.~D.}\ \bibnamefont {Baczewski}}, \ and\ \bibinfo {author}
  {\bibfnamefont {S.}~\bibnamefont {Misra}},\ }\href@noop {} {\bibfield
  {journal} {\bibinfo  {journal} {Journal of Physical Chemistry C}\ }\textbf
  {\bibinfo {volume} {125}},\ \bibinfo {pages} {481} (\bibinfo {year}
  {2021})}\BibitemShut {NoStop}%
\bibitem [{\citenamefont {Herner}\ and\ \citenamefont
  {Clark}(2004)}]{Herner2004}%
  \BibitemOpen
  \bibfield  {author} {\bibinfo {author} {\bibfnamefont {S.~B.}\ \bibnamefont
  {Herner}}\ and\ \bibinfo {author} {\bibfnamefont {M.~H.}\ \bibnamefont
  {Clark}},\ }\href@noop {} {\bibfield  {journal} {\bibinfo  {journal} {Journal
  of Vacuum Science {\&} Technology B: Microelectronics and Nanometer
  Structures}\ }\textbf {\bibinfo {volume} {22}},\ \bibinfo {pages} {1}
  (\bibinfo {year} {2004})}\BibitemShut {NoStop}%
\bibitem [{\citenamefont {Consiglio}\ \emph {et~al.}(2016)\citenamefont
  {Consiglio}, \citenamefont {Clark}, \citenamefont {O'Meara}, \citenamefont
  {Wajda}, \citenamefont {Tapily},\ and\ \citenamefont
  {Leusink}}]{Consiglio2016}%
  \BibitemOpen
  \bibfield  {author} {\bibinfo {author} {\bibfnamefont {S.}~\bibnamefont
  {Consiglio}}, \bibinfo {author} {\bibfnamefont {R.~D.}\ \bibnamefont
  {Clark}}, \bibinfo {author} {\bibfnamefont {D.}~\bibnamefont {O'Meara}},
  \bibinfo {author} {\bibfnamefont {C.~S.}\ \bibnamefont {Wajda}}, \bibinfo
  {author} {\bibfnamefont {K.}~\bibnamefont {Tapily}}, \ and\ \bibinfo {author}
  {\bibfnamefont {G.~J.}\ \bibnamefont {Leusink}},\ }\href@noop {} {\bibfield
  {journal} {\bibinfo  {journal} {Journal of Vacuum Science {\&} Technology A:
  Vacuum, Surfaces, and Films}\ }\textbf {\bibinfo {volume} {34}},\ \bibinfo
  {pages} {01A102} (\bibinfo {year} {2016})}\BibitemShut {NoStop}%
\bibitem [{\citenamefont {Pilli}\ \emph {et~al.}(2018)\citenamefont {Pilli},
  \citenamefont {Jones}, \citenamefont {Lee}, \citenamefont {Chugh},
  \citenamefont {Kelber}, \citenamefont {Pasquale},\ and\ \citenamefont
  {LaVoie}}]{Pilli:2018}%
  \BibitemOpen
  \bibfield  {author} {\bibinfo {author} {\bibfnamefont {A.}~\bibnamefont
  {Pilli}}, \bibinfo {author} {\bibfnamefont {J.}~\bibnamefont {Jones}},
  \bibinfo {author} {\bibfnamefont {V.}~\bibnamefont {Lee}}, \bibinfo {author}
  {\bibfnamefont {N.}~\bibnamefont {Chugh}}, \bibinfo {author} {\bibfnamefont
  {J.}~\bibnamefont {Kelber}}, \bibinfo {author} {\bibfnamefont
  {F.}~\bibnamefont {Pasquale}}, \ and\ \bibinfo {author} {\bibfnamefont
  {A.}~\bibnamefont {LaVoie}},\ }\href {\doibase 10.1116/1.5044396} {\bibfield
  {journal} {\bibinfo  {journal} {Journal of Vacuum Science \& Technology A}\
  }\textbf {\bibinfo {volume} {36}},\ \bibinfo {pages} {061503} (\bibinfo
  {year} {2018})}\BibitemShut {NoStop}%
\bibitem [{\citenamefont {Duvauchelle}\ \emph {et~al.}(2015)\citenamefont
  {Duvauchelle}, \citenamefont {Francheteau}, \citenamefont {Marcenat},
  \citenamefont {Chiodi}, \citenamefont {D{\'{e}}barre}, \citenamefont
  {Hasselbach}, \citenamefont {Kirtley},\ and\ \citenamefont
  {Lefloch}}]{Duvauchelle2015}%
  \BibitemOpen
  \bibfield  {author} {\bibinfo {author} {\bibfnamefont {J.~E.}\ \bibnamefont
  {Duvauchelle}}, \bibinfo {author} {\bibfnamefont {A.}~\bibnamefont
  {Francheteau}}, \bibinfo {author} {\bibfnamefont {C.}~\bibnamefont
  {Marcenat}}, \bibinfo {author} {\bibfnamefont {F.}~\bibnamefont {Chiodi}},
  \bibinfo {author} {\bibfnamefont {D.}~\bibnamefont {D{\'{e}}barre}}, \bibinfo
  {author} {\bibfnamefont {K.}~\bibnamefont {Hasselbach}}, \bibinfo {author}
  {\bibfnamefont {J.~R.}\ \bibnamefont {Kirtley}}, \ and\ \bibinfo {author}
  {\bibfnamefont {F.}~\bibnamefont {Lefloch}},\ }\href@noop {} {\bibfield
  {journal} {\bibinfo  {journal} {Applied Physics Letters}\ }\textbf {\bibinfo
  {volume} {107}},\ \bibinfo {pages} {072601} (\bibinfo {year}
  {2015})}\BibitemShut {NoStop}%
\bibitem [{\citenamefont {Cammilleri}\ \emph {et~al.}(2008)\citenamefont
  {Cammilleri}, \citenamefont {Fossard}, \citenamefont {Débarre},
  \citenamefont {Manh}, \citenamefont {Dubois}, \citenamefont {Bustarret},
  \citenamefont {Marcenat}, \citenamefont {Achatz}, \citenamefont {Bouchier},\
  and\ \citenamefont {Boulmer}}]{CAMMILLERI200875}%
  \BibitemOpen
  \bibfield  {author} {\bibinfo {author} {\bibfnamefont {D.}~\bibnamefont
  {Cammilleri}}, \bibinfo {author} {\bibfnamefont {F.}~\bibnamefont {Fossard}},
  \bibinfo {author} {\bibfnamefont {D.}~\bibnamefont {Débarre}}, \bibinfo
  {author} {\bibfnamefont {C.~T.}\ \bibnamefont {Manh}}, \bibinfo {author}
  {\bibfnamefont {C.}~\bibnamefont {Dubois}}, \bibinfo {author} {\bibfnamefont
  {E.}~\bibnamefont {Bustarret}}, \bibinfo {author} {\bibfnamefont
  {C.}~\bibnamefont {Marcenat}}, \bibinfo {author} {\bibfnamefont
  {P.}~\bibnamefont {Achatz}}, \bibinfo {author} {\bibfnamefont
  {D.}~\bibnamefont {Bouchier}}, \ and\ \bibinfo {author} {\bibfnamefont
  {J.}~\bibnamefont {Boulmer}},\ }\href {\doibase
  https://doi.org/10.1016/j.tsf.2008.08.073} {\bibfield  {journal} {\bibinfo
  {journal} {Thin Solid Films}\ }\textbf {\bibinfo {volume} {517}},\ \bibinfo
  {pages} {75 } (\bibinfo {year} {2008})}\BibitemShut {NoStop}%
\bibitem [{\citenamefont {Trenhaile}\ \emph {et~al.}(2006)\citenamefont
  {Trenhaile}, \citenamefont {Agrawal},\ and\ \citenamefont
  {Weaver}}]{Trenhaile_O:2006}%
  \BibitemOpen
  \bibfield  {author} {\bibinfo {author} {\bibfnamefont {B.~R.}\ \bibnamefont
  {Trenhaile}}, \bibinfo {author} {\bibfnamefont {A.}~\bibnamefont {Agrawal}},
  \ and\ \bibinfo {author} {\bibfnamefont {J.~H.}\ \bibnamefont {Weaver}},\
  }\href {\doibase 10.1063/1.2362623} {\bibfield  {journal} {\bibinfo
  {journal} {Appl. Phys. Lett.}\ }\textbf {\bibinfo {volume} {89}},\ \bibinfo
  {pages} {151917} (\bibinfo {year} {2006})}\BibitemShut {NoStop}%
\bibitem [{\citenamefont {Yu}\ \emph {et~al.}(2008)\citenamefont {Yu},
  \citenamefont {Kim},\ and\ \citenamefont {Koo}}]{Yu:2008}%
  \BibitemOpen
  \bibfield  {author} {\bibinfo {author} {\bibfnamefont {S.-Y.}\ \bibnamefont
  {Yu}}, \bibinfo {author} {\bibfnamefont {H.}~\bibnamefont {Kim}}, \ and\
  \bibinfo {author} {\bibfnamefont {J.-Y.}\ \bibnamefont {Koo}},\ }\href
  {\doibase 10.1103/PhysRevLett.100.036107} {\bibfield  {journal} {\bibinfo
  {journal} {Phys. Rev. Lett.}\ }\textbf {\bibinfo {volume} {100}},\ \bibinfo
  {pages} {036107} (\bibinfo {year} {2008})}\BibitemShut {NoStop}%
\bibitem [{\citenamefont {Ferguson}\ \emph {et~al.}(2009)\citenamefont
  {Ferguson}, \citenamefont {Das},\ and\ \citenamefont
  {Raghavachari}}]{Ferguson:2009}%
  \BibitemOpen
  \bibfield  {author} {\bibinfo {author} {\bibfnamefont {G.~A.}\ \bibnamefont
  {Ferguson}}, \bibinfo {author} {\bibfnamefont {U.}~\bibnamefont {Das}}, \
  and\ \bibinfo {author} {\bibfnamefont {K.}~\bibnamefont {Raghavachari}},\
  }\href {\doibase doi: 10.1021/jp902313d} {\bibfield  {journal} {\bibinfo
  {journal} {J. Phys. Chem. C}\ }\textbf {\bibinfo {volume} {113}},\ \bibinfo
  {pages} {10146} (\bibinfo {year} {2009})}\BibitemShut {NoStop}%
\bibitem [{\citenamefont {Deutsch}\ \emph {et~al.}(1981)\citenamefont
  {Deutsch}, \citenamefont {Ehrlich}, \citenamefont {Rathman}, \citenamefont
  {Silversmith},\ and\ \citenamefont {Osgood~Jr}}]{deutsch1981electrical}%
  \BibitemOpen
  \bibfield  {author} {\bibinfo {author} {\bibfnamefont {T.}~\bibnamefont
  {Deutsch}}, \bibinfo {author} {\bibfnamefont {D.}~\bibnamefont {Ehrlich}},
  \bibinfo {author} {\bibfnamefont {D.}~\bibnamefont {Rathman}}, \bibinfo
  {author} {\bibfnamefont {D.}~\bibnamefont {Silversmith}}, \ and\ \bibinfo
  {author} {\bibfnamefont {R.}~\bibnamefont {Osgood~Jr}},\ }\href@noop {}
  {\bibfield  {journal} {\bibinfo  {journal} {Applied Physics Letters}\
  }\textbf {\bibinfo {volume} {39}},\ \bibinfo {pages} {825} (\bibinfo {year}
  {1981})}\BibitemShut {NoStop}%
\bibitem [{\citenamefont {Kunii}\ \emph {et~al.}(2004)\citenamefont {Kunii},
  \citenamefont {Inokuchi}, \citenamefont {Moriya}, \citenamefont {Kurokawa},\
  and\ \citenamefont {Murota}}]{kunii2004situ}%
  \BibitemOpen
  \bibfield  {author} {\bibinfo {author} {\bibfnamefont {Y.}~\bibnamefont
  {Kunii}}, \bibinfo {author} {\bibfnamefont {Y.}~\bibnamefont {Inokuchi}},
  \bibinfo {author} {\bibfnamefont {A.}~\bibnamefont {Moriya}}, \bibinfo
  {author} {\bibfnamefont {H.}~\bibnamefont {Kurokawa}}, \ and\ \bibinfo
  {author} {\bibfnamefont {J.}~\bibnamefont {Murota}},\ }\href@noop {}
  {\bibfield  {journal} {\bibinfo  {journal} {Applied surface science}\
  }\textbf {\bibinfo {volume} {224}},\ \bibinfo {pages} {68} (\bibinfo {year}
  {2004})}\BibitemShut {NoStop}%
\bibitem [{\citenamefont {Lapiano-Smith}\ and\ \citenamefont
  {McFeely}(1992)}]{lapiano1992chemisorption}%
  \BibitemOpen
  \bibfield  {author} {\bibinfo {author} {\bibfnamefont {D.}~\bibnamefont
  {Lapiano-Smith}}\ and\ \bibinfo {author} {\bibfnamefont {F.}~\bibnamefont
  {McFeely}},\ }\href@noop {} {\bibfield  {journal} {\bibinfo  {journal}
  {Journal of applied physics}\ }\textbf {\bibinfo {volume} {72}},\ \bibinfo
  {pages} {4907} (\bibinfo {year} {1992})}\BibitemShut {NoStop}%
\bibitem [{\citenamefont {Ferguson}\ \emph {et~al.}(2002)\citenamefont
  {Ferguson}, \citenamefont {Weimer},\ and\ \citenamefont
  {George}}]{Ferguson2002}%
  \BibitemOpen
  \bibfield  {author} {\bibinfo {author} {\bibfnamefont {J.~D.}\ \bibnamefont
  {Ferguson}}, \bibinfo {author} {\bibfnamefont {A.~W.}\ \bibnamefont
  {Weimer}}, \ and\ \bibinfo {author} {\bibfnamefont {S.~M.}\ \bibnamefont
  {George}},\ }\href {\doibase 10.1016/S0040-6090(02)00431-5} {\bibfield
  {journal} {\bibinfo  {journal} {Thin Solid Films}\ }\textbf {\bibinfo
  {volume} {413}},\ \bibinfo {pages} {16} (\bibinfo {year} {2002})}\BibitemShut
  {NoStop}%
\bibitem [{\citenamefont {Radue}\ \emph {et~al.}(2021)\citenamefont {Radue},
  \citenamefont {Baek}, \citenamefont {Farzaneh}, \citenamefont {Dwyer},
  \citenamefont {Campbell}, \citenamefont {Baczewski}, \citenamefont
  {Bussmann}, \citenamefont {Wang}, \citenamefont {Mo}, \citenamefont {Misra},\
  and\ \citenamefont {Butera}}]{Radue2021}%
  \BibitemOpen
  \bibfield  {author} {\bibinfo {author} {\bibfnamefont {M.~S.}\ \bibnamefont
  {Radue}}, \bibinfo {author} {\bibfnamefont {S.}~\bibnamefont {Baek}},
  \bibinfo {author} {\bibfnamefont {A.}~\bibnamefont {Farzaneh}}, \bibinfo
  {author} {\bibfnamefont {K.~J.}\ \bibnamefont {Dwyer}}, \bibinfo {author}
  {\bibfnamefont {Q.}~\bibnamefont {Campbell}}, \bibinfo {author}
  {\bibfnamefont {A.~D.}\ \bibnamefont {Baczewski}}, \bibinfo {author}
  {\bibfnamefont {E.}~\bibnamefont {Bussmann}}, \bibinfo {author}
  {\bibfnamefont {G.~T.}\ \bibnamefont {Wang}}, \bibinfo {author}
  {\bibfnamefont {Y.}~\bibnamefont {Mo}}, \bibinfo {author} {\bibfnamefont
  {S.}~\bibnamefont {Misra}}, \ and\ \bibinfo {author} {\bibfnamefont {R.~E.}\
  \bibnamefont {Butera}},\ }\href {http://arxiv.org/abs/2101.09265} {\
  (\bibinfo {year} {2021})},\ \Eprint {http://arxiv.org/abs/2101.09265}
  {arXiv:2101.09265} \BibitemShut {NoStop}%
\bibitem [{\citenamefont {Liu}\ \emph {et~al.}(2008)\citenamefont {Liu},
  \citenamefont {Zhang},\ and\ \citenamefont {Zhu}}]{Liu2008}%
  \BibitemOpen
  \bibfield  {author} {\bibinfo {author} {\bibfnamefont {Z.}~\bibnamefont
  {Liu}}, \bibinfo {author} {\bibfnamefont {Z.}~\bibnamefont {Zhang}}, \ and\
  \bibinfo {author} {\bibfnamefont {X.}~\bibnamefont {Zhu}},\ }\href@noop {}
  {\bibfield  {journal} {\bibinfo  {journal} {Physical Review B - Condensed
  Matter and Materials Physics}\ }\textbf {\bibinfo {volume} {77}},\ \bibinfo
  {pages} {035322} (\bibinfo {year} {2008})}\BibitemShut {NoStop}%
\bibitem [{\citenamefont {Wang}\ and\ \citenamefont {Hamers}(1995)}]{Wang1995}%
  \BibitemOpen
  \bibfield  {author} {\bibinfo {author} {\bibfnamefont {Y.}~\bibnamefont
  {Wang}}\ and\ \bibinfo {author} {\bibfnamefont {R.~J.}\ \bibnamefont
  {Hamers}},\ }\href@noop {} {\bibfield  {journal} {\bibinfo  {journal}
  {Applied Physics Letters}\ }\textbf {\bibinfo {volume} {66}},\ \bibinfo
  {pages} {2057} (\bibinfo {year} {1995})}\BibitemShut {NoStop}%
\bibitem [{\citenamefont {Gautier}\ \emph {et~al.}(1996)\citenamefont
  {Gautier}, \citenamefont {Prost}, \citenamefont {Prudon},\ and\ \citenamefont
  {Dupuy}}]{Gautier1996}%
  \BibitemOpen
  \bibfield  {author} {\bibinfo {author} {\bibfnamefont {B.}~\bibnamefont
  {Gautier}}, \bibinfo {author} {\bibfnamefont {R.}~\bibnamefont {Prost}},
  \bibinfo {author} {\bibfnamefont {G.}~\bibnamefont {Prudon}}, \ and\ \bibinfo
  {author} {\bibfnamefont {J.~C.}\ \bibnamefont {Dupuy}},\ }\href@noop {}
  {\bibfield  {journal} {\bibinfo  {journal} {Surface and Interface Analysis}\
  }\textbf {\bibinfo {volume} {24}},\ \bibinfo {pages} {733} (\bibinfo {year}
  {1996})}\BibitemShut {NoStop}%
\bibitem [{\citenamefont {Hagmann}\ \emph {et~al.}(2018)\citenamefont
  {Hagmann}, \citenamefont {Wang}, \citenamefont {Namboodiri}, \citenamefont
  {Wyrick}, \citenamefont {Murray}, \citenamefont {Stewart}, \citenamefont
  {Silver},\ and\ \citenamefont {Richter}}]{Hagmann2018}%
  \BibitemOpen
  \bibfield  {author} {\bibinfo {author} {\bibfnamefont {J.~A.}\ \bibnamefont
  {Hagmann}}, \bibinfo {author} {\bibfnamefont {X.}~\bibnamefont {Wang}},
  \bibinfo {author} {\bibfnamefont {P.}~\bibnamefont {Namboodiri}}, \bibinfo
  {author} {\bibfnamefont {J.}~\bibnamefont {Wyrick}}, \bibinfo {author}
  {\bibfnamefont {R.}~\bibnamefont {Murray}}, \bibinfo {author} {\bibfnamefont
  {M.~D.}\ \bibnamefont {Stewart}}, \bibinfo {author} {\bibfnamefont {R.~M.}\
  \bibnamefont {Silver}}, \ and\ \bibinfo {author} {\bibfnamefont {C.~A.}\
  \bibnamefont {Richter}},\ }\href@noop {} {\bibfield  {journal} {\bibinfo
  {journal} {Applied Physics Letters}\ }\textbf {\bibinfo {volume} {112}},\
  \bibinfo {pages} {043102} (\bibinfo {year} {2018})}\BibitemShut {NoStop}%
\bibitem [{\citenamefont {Weir}\ \emph {et~al.}(1994)\citenamefont {Weir},
  \citenamefont {Feldman}, \citenamefont {Monroe}, \citenamefont {Grossmann},
  \citenamefont {Headrick},\ and\ \citenamefont {Hart}}]{Weir1994}%
  \BibitemOpen
  \bibfield  {author} {\bibinfo {author} {\bibfnamefont {B.~E.}\ \bibnamefont
  {Weir}}, \bibinfo {author} {\bibfnamefont {L.~C.}\ \bibnamefont {Feldman}},
  \bibinfo {author} {\bibfnamefont {D.}~\bibnamefont {Monroe}}, \bibinfo
  {author} {\bibfnamefont {H.~J.}\ \bibnamefont {Grossmann}}, \bibinfo {author}
  {\bibfnamefont {R.~L.}\ \bibnamefont {Headrick}}, \ and\ \bibinfo {author}
  {\bibfnamefont {T.~R.}\ \bibnamefont {Hart}},\ }\href@noop {} {\bibfield
  {journal} {\bibinfo  {journal} {Applied Physics Letters}\ }\textbf {\bibinfo
  {volume} {65}},\ \bibinfo {pages} {737} (\bibinfo {year} {1994})}\BibitemShut
  {NoStop}%
\bibitem [{\citenamefont {Hagmann}\ \emph {et~al.}(2020)\citenamefont
  {Hagmann}, \citenamefont {Wang}, \citenamefont {Kashid}, \citenamefont
  {Namboodiri}, \citenamefont {Wyrick}, \citenamefont {Schmucker},
  \citenamefont {Stewart}, \citenamefont {Silver},\ and\ \citenamefont
  {Richter}}]{Hagmann2020}%
  \BibitemOpen
  \bibfield  {author} {\bibinfo {author} {\bibfnamefont {J.~A.}\ \bibnamefont
  {Hagmann}}, \bibinfo {author} {\bibfnamefont {X.}~\bibnamefont {Wang}},
  \bibinfo {author} {\bibfnamefont {R.}~\bibnamefont {Kashid}}, \bibinfo
  {author} {\bibfnamefont {P.}~\bibnamefont {Namboodiri}}, \bibinfo {author}
  {\bibfnamefont {J.}~\bibnamefont {Wyrick}}, \bibinfo {author} {\bibfnamefont
  {S.~W.}\ \bibnamefont {Schmucker}}, \bibinfo {author} {\bibfnamefont {M.~D.}\
  \bibnamefont {Stewart}}, \bibinfo {author} {\bibfnamefont {R.~M.}\
  \bibnamefont {Silver}}, \ and\ \bibinfo {author} {\bibfnamefont {C.~A.}\
  \bibnamefont {Richter}},\ }\href@noop {} {\bibfield  {journal} {\bibinfo
  {journal} {Physical Review B}\ }\textbf {\bibinfo {volume} {101}},\ \bibinfo
  {pages} {245419} (\bibinfo {year} {2020})}\BibitemShut {NoStop}%
\bibitem [{\citenamefont {Pavlova}\ \emph {et~al.}(2019)\citenamefont
  {Pavlova}, \citenamefont {Skorokhodov}, \citenamefont {Zhidomirov},\ and\
  \citenamefont {Eltsov}}]{pavlova2019ab}%
  \BibitemOpen
  \bibfield  {author} {\bibinfo {author} {\bibfnamefont {T.~V.}\ \bibnamefont
  {Pavlova}}, \bibinfo {author} {\bibfnamefont {E.~S.}\ \bibnamefont
  {Skorokhodov}}, \bibinfo {author} {\bibfnamefont {G.~M.}\ \bibnamefont
  {Zhidomirov}}, \ and\ \bibinfo {author} {\bibfnamefont {K.~N.}\ \bibnamefont
  {Eltsov}},\ }\href@noop {} {\bibfield  {journal} {\bibinfo  {journal} {The
  Journal of Physical Chemistry C}\ }\textbf {\bibinfo {volume} {123}},\
  \bibinfo {pages} {19806} (\bibinfo {year} {2019})}\BibitemShut {NoStop}%
\bibitem [{\citenamefont {Zhang}\ and\ \citenamefont
  {Lagally}(1994)}]{zhang1994atomic}%
  \BibitemOpen
  \bibfield  {author} {\bibinfo {author} {\bibfnamefont {Z.}~\bibnamefont
  {Zhang}}\ and\ \bibinfo {author} {\bibfnamefont {M.~G.}\ \bibnamefont
  {Lagally}},\ }\href@noop {} {\bibfield  {journal} {\bibinfo  {journal}
  {Physical review letters}\ }\textbf {\bibinfo {volume} {72}},\ \bibinfo
  {pages} {693} (\bibinfo {year} {1994})}\BibitemShut {NoStop}%
\bibitem [{\citenamefont {Voigtl{\"a}nder}\ and\ \citenamefont
  {Zinner}(1994)}]{voigtlander1994surfactant}%
  \BibitemOpen
  \bibfield  {author} {\bibinfo {author} {\bibfnamefont {B.}~\bibnamefont
  {Voigtl{\"a}nder}}\ and\ \bibinfo {author} {\bibfnamefont {A.}~\bibnamefont
  {Zinner}},\ }\href@noop {} {\bibfield  {journal} {\bibinfo  {journal}
  {Journal of Vacuum Science \& Technology A: Vacuum, Surfaces, and Films}\
  }\textbf {\bibinfo {volume} {12}},\ \bibinfo {pages} {1932} (\bibinfo {year}
  {1994})}\BibitemShut {NoStop}%
\bibitem [{\citenamefont {Ferng}\ \emph {et~al.}(2009)\citenamefont {Ferng},
  \citenamefont {Wu}, \citenamefont {Lin},\ and\ \citenamefont
  {Chiang}}]{Ferng2009}%
  \BibitemOpen
  \bibfield  {author} {\bibinfo {author} {\bibfnamefont {S.-S.}\ \bibnamefont
  {Ferng}}, \bibinfo {author} {\bibfnamefont {S.-T.}\ \bibnamefont {Wu}},
  \bibinfo {author} {\bibfnamefont {D.-S.}\ \bibnamefont {Lin}}, \ and\
  \bibinfo {author} {\bibfnamefont {T.~C.}\ \bibnamefont {Chiang}},\ }\href
  {\doibase 10.1063/1.3122987} {\bibfield  {journal} {\bibinfo  {journal} {J.
  Chem. Phys.}\ }\textbf {\bibinfo {volume} {130}},\ \bibinfo {pages} {164706}
  (\bibinfo {year} {2009})}\BibitemShut {NoStop}%
\bibitem [{\citenamefont {Li}\ \emph {et~al.}(2011)\citenamefont {Li},
  \citenamefont {Chang}, \citenamefont {Chien}, \citenamefont {Chang},
  \citenamefont {Chiang},\ and\ \citenamefont {Lin}}]{Li:2011}%
  \BibitemOpen
  \bibfield  {author} {\bibinfo {author} {\bibfnamefont {H.-D.}\ \bibnamefont
  {Li}}, \bibinfo {author} {\bibfnamefont {C.-Y.}\ \bibnamefont {Chang}},
  \bibinfo {author} {\bibfnamefont {L.-Y.}\ \bibnamefont {Chien}}, \bibinfo
  {author} {\bibfnamefont {S.-H.}\ \bibnamefont {Chang}}, \bibinfo {author}
  {\bibfnamefont {T.-C.}\ \bibnamefont {Chiang}}, \ and\ \bibinfo {author}
  {\bibfnamefont {D.-S.}\ \bibnamefont {Lin}},\ }\href {\doibase
  10.1103/PhysRevB.83.075403} {\bibfield  {journal} {\bibinfo  {journal} {Phys.
  Rev. B}\ }\textbf {\bibinfo {volume} {83}},\ \bibinfo {pages} {075403}
  (\bibinfo {year} {2011})}\BibitemShut {NoStop}%
\bibitem [{\citenamefont {Mitsui}\ \emph {et~al.}(1999)\citenamefont {Mitsui},
  \citenamefont {Curtis},\ and\ \citenamefont {Ganz}}]{Mitsui:1999}%
  \BibitemOpen
  \bibfield  {author} {\bibinfo {author} {\bibfnamefont {T.}~\bibnamefont
  {Mitsui}}, \bibinfo {author} {\bibfnamefont {R.}~\bibnamefont {Curtis}}, \
  and\ \bibinfo {author} {\bibfnamefont {E.}~\bibnamefont {Ganz}},\ }\href
  {\doibase 10.1063/1.370946} {\bibfield  {journal} {\bibinfo  {journal} {J.
  Appl. Phys.}\ }\textbf {\bibinfo {volume} {86}},\ \bibinfo {pages} {1676}
  (\bibinfo {year} {1999})}\BibitemShut {NoStop}%
\bibitem [{\citenamefont {Dick}\ \emph {et~al.}(2016)\citenamefont {Dick},
  \citenamefont {Ballard}, \citenamefont {Longo}, \citenamefont {Randall},
  \citenamefont {Cho},\ and\ \citenamefont {Chabal}}]{TiCl4:Zyvex}%
  \BibitemOpen
  \bibfield  {author} {\bibinfo {author} {\bibfnamefont {D.}~\bibnamefont
  {Dick}}, \bibinfo {author} {\bibfnamefont {J.~B.}\ \bibnamefont {Ballard}},
  \bibinfo {author} {\bibfnamefont {R.~C.}\ \bibnamefont {Longo}}, \bibinfo
  {author} {\bibfnamefont {J.~N.}\ \bibnamefont {Randall}}, \bibinfo {author}
  {\bibfnamefont {K.}~\bibnamefont {Cho}}, \ and\ \bibinfo {author}
  {\bibfnamefont {Y.~J.}\ \bibnamefont {Chabal}},\ }\href {\doibase
  10.1021/acs.jpcc.6b08130} {\bibfield  {journal} {\bibinfo  {journal} {The
  Journal of Physical Chemistry C}\ }\textbf {\bibinfo {volume} {120}},\
  \bibinfo {pages} {24213} (\bibinfo {year} {2016})},\ \Eprint
  {http://arxiv.org/abs/https://doi.org/10.1021/acs.jpcc.6b08130}
  {https://doi.org/10.1021/acs.jpcc.6b08130} \BibitemShut {NoStop}%
\bibitem [{\citenamefont {Longo}\ \emph {et~al.}(2014)\citenamefont {Longo},
  \citenamefont {McDonnell}, \citenamefont {Dick}, \citenamefont {Wallace},
  \citenamefont {Chabal}, \citenamefont {Owen}, \citenamefont {Ballard},
  \citenamefont {Randall},\ and\ \citenamefont {Cho}}]{TiCl4:Zyvex2}%
  \BibitemOpen
  \bibfield  {author} {\bibinfo {author} {\bibfnamefont {R.~C.}\ \bibnamefont
  {Longo}}, \bibinfo {author} {\bibfnamefont {S.}~\bibnamefont {McDonnell}},
  \bibinfo {author} {\bibfnamefont {D.}~\bibnamefont {Dick}}, \bibinfo {author}
  {\bibfnamefont {R.~M.}\ \bibnamefont {Wallace}}, \bibinfo {author}
  {\bibfnamefont {Y.~J.}\ \bibnamefont {Chabal}}, \bibinfo {author}
  {\bibfnamefont {J.~H.~G.}\ \bibnamefont {Owen}}, \bibinfo {author}
  {\bibfnamefont {J.~B.}\ \bibnamefont {Ballard}}, \bibinfo {author}
  {\bibfnamefont {J.~N.}\ \bibnamefont {Randall}}, \ and\ \bibinfo {author}
  {\bibfnamefont {K.}~\bibnamefont {Cho}},\ }\href {\doibase 10.1116/1.4864619}
  {\bibfield  {journal} {\bibinfo  {journal} {Journal of Vacuum Science \&
  Technology B}\ }\textbf {\bibinfo {volume} {32}},\ \bibinfo {pages} {03D112}
  (\bibinfo {year} {2014})},\ \Eprint
  {http://arxiv.org/abs/https://doi.org/10.1116/1.4864619}
  {https://doi.org/10.1116/1.4864619} \BibitemShut {NoStop}%
\bibitem [{\citenamefont {Pavlova}\ \emph {et~al.}(2018)\citenamefont
  {Pavlova}, \citenamefont {Zhidomirov},\ and\ \citenamefont
  {Eltsov}}]{Pavlova:2018}%
  \BibitemOpen
  \bibfield  {author} {\bibinfo {author} {\bibfnamefont {T.~V.}\ \bibnamefont
  {Pavlova}}, \bibinfo {author} {\bibfnamefont {G.~M.}\ \bibnamefont
  {Zhidomirov}}, \ and\ \bibinfo {author} {\bibfnamefont {K.~N.}\ \bibnamefont
  {Eltsov}},\ }\href {\doibase 10.1021/acs.jpcc.7b11519} {\bibfield  {journal}
  {\bibinfo  {journal} {J. Phys. Chem. C}\ }\textbf {\bibinfo {volume} {122}},\
  \bibinfo {pages} {1741} (\bibinfo {year} {2018})}\BibitemShut {NoStop}%
\bibitem [{\citenamefont {Dwyer}\ \emph {et~al.}(2019)\citenamefont {Dwyer},
  \citenamefont {Dreyer},\ and\ \citenamefont {Butera}}]{ClLitho}%
  \BibitemOpen
  \bibfield  {author} {\bibinfo {author} {\bibfnamefont {K.~J.}\ \bibnamefont
  {Dwyer}}, \bibinfo {author} {\bibfnamefont {M.}~\bibnamefont {Dreyer}}, \
  and\ \bibinfo {author} {\bibfnamefont {R.~E.}\ \bibnamefont {Butera}},\
  }\href {\doibase 10.1021/acs.jpca.9b07127} {\bibfield  {journal} {\bibinfo
  {journal} {J. Phys. Chem. A}\ }\textbf {\bibinfo {volume} {123}},\ \bibinfo
  {pages} {10793} (\bibinfo {year} {2019})}\BibitemShut {NoStop}%
\bibitem [{\citenamefont {Silva-Quinones}\ \emph {et~al.}(2020)\citenamefont
  {Silva-Quinones}, \citenamefont {He}, \citenamefont {Butera}, \citenamefont
  {Wang},\ and\ \citenamefont {Teplyakov}}]{silva2020reaction}%
  \BibitemOpen
  \bibfield  {author} {\bibinfo {author} {\bibfnamefont {D.}~\bibnamefont
  {Silva-Quinones}}, \bibinfo {author} {\bibfnamefont {C.}~\bibnamefont {He}},
  \bibinfo {author} {\bibfnamefont {R.~E.}\ \bibnamefont {Butera}}, \bibinfo
  {author} {\bibfnamefont {G.~T.}\ \bibnamefont {Wang}}, \ and\ \bibinfo
  {author} {\bibfnamefont {A.~V.}\ \bibnamefont {Teplyakov}},\ }\href@noop {}
  {\bibfield  {journal} {\bibinfo  {journal} {Applied Surface Science}\ ,\
  \bibinfo {pages} {146907}} (\bibinfo {year} {2020})}\BibitemShut {NoStop}%
\bibitem [{\citenamefont {Zhang}\ \emph {et~al.}(1996)\citenamefont {Zhang},
  \citenamefont {Kulakov}, \citenamefont {Bullemer}, \citenamefont {Eisele},\
  and\ \citenamefont {Zotov}}]{Zhang1996}%
  \BibitemOpen
  \bibfield  {author} {\bibinfo {author} {\bibfnamefont {Z.}~\bibnamefont
  {Zhang}}, \bibinfo {author} {\bibfnamefont {M.~A.}\ \bibnamefont {Kulakov}},
  \bibinfo {author} {\bibfnamefont {B.}~\bibnamefont {Bullemer}}, \bibinfo
  {author} {\bibfnamefont {I.}~\bibnamefont {Eisele}}, \ and\ \bibinfo {author}
  {\bibfnamefont {A.~V.}\ \bibnamefont {Zotov}},\ }\href@noop {} {\bibfield
  {journal} {\bibinfo  {journal} {Applied Physics Letters}\ }\textbf {\bibinfo
  {volume} {69}},\ \bibinfo {pages} {494} (\bibinfo {year} {1996})}\BibitemShut
  {NoStop}%
\bibitem [{\citenamefont {Mackus}\ \emph {et~al.}(2019)\citenamefont {Mackus},
  \citenamefont {Merkx},\ and\ \citenamefont {Kessels}}]{Mackus2019}%
  \BibitemOpen
  \bibfield  {author} {\bibinfo {author} {\bibfnamefont {A.~J.}\ \bibnamefont
  {Mackus}}, \bibinfo {author} {\bibfnamefont {M.~J.}\ \bibnamefont {Merkx}}, \
  and\ \bibinfo {author} {\bibfnamefont {W.~M.}\ \bibnamefont {Kessels}},\
  }\href {\doibase 10.1021/acs.chemmater.8b03454} {\bibfield  {journal}
  {\bibinfo  {journal} {Chemistry of Materials}\ }\textbf {\bibinfo {volume}
  {31}},\ \bibinfo {pages} {2} (\bibinfo {year} {2019})}\BibitemShut {NoStop}%
\bibitem [{\citenamefont {Gladfelter}(1993)}]{Gladfelter1993}%
  \BibitemOpen
  \bibfield  {author} {\bibinfo {author} {\bibfnamefont {W.~L.}\ \bibnamefont
  {Gladfelter}},\ }\href {\doibase 10.1021/cm00034a004} {\bibfield  {journal}
  {\bibinfo  {journal} {Chemistry of Materials}\ }\textbf {\bibinfo {volume}
  {5}},\ \bibinfo {pages} {1372} (\bibinfo {year} {1993})}\BibitemShut
  {NoStop}%
\bibitem [{\citenamefont {Ramanayaka}\ \emph {et~al.}(2018)\citenamefont
  {Ramanayaka}, \citenamefont {Kim}, \citenamefont {Hagmann}, \citenamefont
  {Murray}, \citenamefont {Tang}, \citenamefont {Meisenkothen}, \citenamefont
  {Zhang}, \citenamefont {Bendersky}, \citenamefont {Davydov}, \citenamefont
  {Zimmerman}, \citenamefont {Richter},\ and\ \citenamefont
  {Pomeroy}}]{doi:10.1063/1.5045338}%
  \BibitemOpen
  \bibfield  {author} {\bibinfo {author} {\bibfnamefont {A.~N.}\ \bibnamefont
  {Ramanayaka}}, \bibinfo {author} {\bibfnamefont {H.-S.}\ \bibnamefont {Kim}},
  \bibinfo {author} {\bibfnamefont {J.~A.}\ \bibnamefont {Hagmann}}, \bibinfo
  {author} {\bibfnamefont {R.~E.}\ \bibnamefont {Murray}}, \bibinfo {author}
  {\bibfnamefont {K.}~\bibnamefont {Tang}}, \bibinfo {author} {\bibfnamefont
  {F.}~\bibnamefont {Meisenkothen}}, \bibinfo {author} {\bibfnamefont {H.~R.}\
  \bibnamefont {Zhang}}, \bibinfo {author} {\bibfnamefont {L.~A.}\ \bibnamefont
  {Bendersky}}, \bibinfo {author} {\bibfnamefont {A.~V.}\ \bibnamefont
  {Davydov}}, \bibinfo {author} {\bibfnamefont {N.~M.}\ \bibnamefont
  {Zimmerman}}, \bibinfo {author} {\bibfnamefont {C.~A.}\ \bibnamefont
  {Richter}}, \ and\ \bibinfo {author} {\bibfnamefont {J.~M.}\ \bibnamefont
  {Pomeroy}},\ }\href {\doibase 10.1063/1.5045338} {\bibfield  {journal}
  {\bibinfo  {journal} {AIP Advances}\ }\textbf {\bibinfo {volume} {8}},\
  \bibinfo {pages} {075329} (\bibinfo {year} {2018})},\ \Eprint
  {http://arxiv.org/abs/https://doi.org/10.1063/1.5045338}
  {https://doi.org/10.1063/1.5045338} \BibitemShut {NoStop}%
\bibitem [{\citenamefont {Goh}\ \emph {et~al.}(2006)\citenamefont {Goh},
  \citenamefont {Oberbeck}, \citenamefont {Simmons}, \citenamefont {Hamilton},\
  and\ \citenamefont {Butcher}}]{goh2006influence}%
  \BibitemOpen
  \bibfield  {author} {\bibinfo {author} {\bibfnamefont {K.}~\bibnamefont
  {Goh}}, \bibinfo {author} {\bibfnamefont {L.}~\bibnamefont {Oberbeck}},
  \bibinfo {author} {\bibfnamefont {M.}~\bibnamefont {Simmons}}, \bibinfo
  {author} {\bibfnamefont {A.}~\bibnamefont {Hamilton}}, \ and\ \bibinfo
  {author} {\bibfnamefont {M.}~\bibnamefont {Butcher}},\ }\href@noop {}
  {\bibfield  {journal} {\bibinfo  {journal} {Physical Review B}\ }\textbf
  {\bibinfo {volume} {73}},\ \bibinfo {pages} {035401} (\bibinfo {year}
  {2006})}\BibitemShut {NoStop}%
\bibitem [{\citenamefont {Weber}\ \emph {et~al.}(2012)\citenamefont {Weber},
  \citenamefont {Mahapatra}, \citenamefont {Ryu}, \citenamefont {Lee},
  \citenamefont {Fuhrer}, \citenamefont {Reusch}, \citenamefont {Thompson},
  \citenamefont {Lee}, \citenamefont {Klimeck}, \citenamefont {Hollenberg}
  \emph {et~al.}}]{weber2012ohm}%
  \BibitemOpen
  \bibfield  {author} {\bibinfo {author} {\bibfnamefont {B.}~\bibnamefont
  {Weber}}, \bibinfo {author} {\bibfnamefont {S.}~\bibnamefont {Mahapatra}},
  \bibinfo {author} {\bibfnamefont {H.}~\bibnamefont {Ryu}}, \bibinfo {author}
  {\bibfnamefont {S.}~\bibnamefont {Lee}}, \bibinfo {author} {\bibfnamefont
  {A.}~\bibnamefont {Fuhrer}}, \bibinfo {author} {\bibfnamefont
  {T.}~\bibnamefont {Reusch}}, \bibinfo {author} {\bibfnamefont
  {D.}~\bibnamefont {Thompson}}, \bibinfo {author} {\bibfnamefont
  {W.}~\bibnamefont {Lee}}, \bibinfo {author} {\bibfnamefont {G.}~\bibnamefont
  {Klimeck}}, \bibinfo {author} {\bibfnamefont {L.~C.}\ \bibnamefont
  {Hollenberg}},  \emph {et~al.},\ }\href@noop {} {\bibfield  {journal}
  {\bibinfo  {journal} {Science}\ }\textbf {\bibinfo {volume} {335}},\ \bibinfo
  {pages} {64} (\bibinfo {year} {2012})}\BibitemShut {NoStop}%
\bibitem [{\citenamefont {Lee}\ and\ \citenamefont
  {Ramakrishnan}(1985)}]{lee1985disordered}%
  \BibitemOpen
  \bibfield  {author} {\bibinfo {author} {\bibfnamefont {P.~A.}\ \bibnamefont
  {Lee}}\ and\ \bibinfo {author} {\bibfnamefont {T.}~\bibnamefont
  {Ramakrishnan}},\ }\href@noop {} {\bibfield  {journal} {\bibinfo  {journal}
  {Reviews of Modern Physics}\ }\textbf {\bibinfo {volume} {57}},\ \bibinfo
  {pages} {287} (\bibinfo {year} {1985})}\BibitemShut {NoStop}%
\end{thebibliography}%

\end{document}